\documentclass[twocolumn,showpacs,prb,showkeys]{revtex4}
\usepackage{hyperref}
\usepackage{amsmath,amssymb}
\usepackage{epsfig}
\usepackage{epstopdf}
\usepackage{graphicx}
\usepackage{graphics}
\usepackage{dcolumn}

\begin{document}

\title{Long-range proximity-induced superconductivity in polycrystalline Co nanowires}
\author{M. Kompaniiets$^1$, O. V. Dobrovolskiy$^{1,2}$, C. Neetzel$^3$, E. Begun$^1$, F.~Porrati$^1$, W. Ensinger$^3$ and M. Huth$^1$}
\address
        {$^1$Physikalisches Institut, Goethe-University, 60438 Frankfurt am Main, Germany\\
        $^2$Physical Department, V. Karazin Kharkiv National University, 61077 Kharkiv, Ukraine\\
        $^3$Department of Materials Science, Darmstadt University of Technology, 64287 Darmstadt, Germany}
\date{\today}

\begin{abstract}
We report an experimental study of proximity effect-induced superconductivity in crystalline Cu and Co nanowires and a nanogranular Co nanowire structure in contact with a superconducting W floating electrode which we call inducer. The nanowires were grown by electrochemical deposition in heavy-ion-track etched polycarbonate templates. The nanogranular Co structure was fabricated by focused electron beam induced deposition (FEBID), while the amorphous W inducer was obtained by focused ion beam induced deposition (FIBID). For electrical resistance measurements up to three pairs of  Pt voltage leads were deposited by FIBID at different distances beside the inner inducer electrode, thus allowing us to probe the proximity effect over a length of $2-12~\mu$m. Relative $R(T)$ drops of the same order of magnitude have been observed for the Co and Cu nanowires when sweeping the temperature below 5.2~K ($T_c$ of the FIBID-deposited W inducer). By contrast, relative $R(T)$ drops were found to be an order of magnitude smaller for the nanogranular Co nanowire structure. Our analysis of the resistance data shows that the superconducting proximity length in crystalline Cu and Co is about $1~\mu$m at low temperatures, attesting to a long-range proximity effect in the case of ferromagnetic Co. Moreover, this long-range proximity effect has been revealed to be insusceptible to magnetic fields up to $11$~T, which is indicative of spin-triplet pairing. At the same time, in the nanogranular Co structure proximity-induced superconductivity is strongly suppressed due to the dominating Cooper pair scattering caused by the intrinsic microstructure of the FEBID deposit.
\end{abstract}


\maketitle
\section{Introduction}
If a superconductor~(S) is placed in direct contact with a normal metal~(N), the Cooper pairs penetrate N inducing there superconducting correlations which decay over some distance. This phenomenon is the ``classical'' proximity effect and is comprehensively addressed, e.g., in~\cite{Gen64rmp,Deu69boo}. The characteristic decay distance is called the proximity length and its typical value amounts to $\xi_N \simeq1~\mu$m at low temperatures~\cite{Gen64rmp,Buz05rmp}. Another situation prevails, if N is replaced by a ferromagnet~(F)~\cite{Gen64rmp}. In most superconductors the wave function of the Cooper pairs is singlet as it is formed by two electrons with opposite spins. Theqrefore, if the exchange field $h_{ex}$ in F is homogenous, it tends to align both spins in the same direction. This results in a strong pair-breaking effect and causes a rapid exponential oscillatory decay of the superconducting order parameter in F over a distance $\xi_F$. This effect is short-ranged, with a spin-singlet decay length $\xi_F\simeq1$~nm, as revealed in experiments~\cite{Chi07ltp,Aum01prb}. However, under some circumstances superconductivity is not necessarily suppressed by ferromagnetism as the presence of F may lead to triplet superconducting pairing~\cite{Ber05rmp,Buz05rmp,Esc08nat}. In the triplet state, a Cooper pair is formed by two electrons with parallel spins that makes it insusceptible to the exchange field. As theoretically shown by Bergeret \emph{et al.}~\cite{Ber01prl}, a local inhomogeneity of the magnetization in the vicinity of the S/F interface provides a necessary condition for the spin-triplet pairing in S/F structures. Non-homogeneities of the exchange field can be either intrinsic to F, or arise as a result of experimental manipulations leading to a non-homogeneous alignment of the magnetic moments~\cite{Ber05rmp}. The spin-triplet proximity effect is long-ranged, with a proximity length $\xi_F$ of the same order of magnitude as $\xi_N$~\cite{Ber01prl,Wan10nat}.

In the last decade, the study of the classical proximity effect at an S/N interface as well as the long-range effect at an S/F interface has become a matter of extensive research, both theoretically~\cite{Esc08nat,Lin09prl,Hou07prb} and experimentally~\cite{Kei06nat,Muh96prl,Laz00prb,Gar02prb,Rya01prl,Zdr13prb,Zdr06prl,Wan10nat,Wan09prl,Liu12pcs}. In particular, theoretical works have largely been focused on clarifying the role of local magnetic inhomogeneity near an S/F interface~\cite{Ber01prl,Esc08nat}, others dealt with developing new types of spin-valves based on S/F multilayers~\cite{Tag99prl}, and studying new types of Josephson junctions based on S/F/S trilayers~\cite{Vol03prl,Hou07prb}. Experimentally, to elaborate these and other related problems, most of the studies utilized (multi-) sandwich heterostructures of flat films~\cite{Kei06nat,Muh96prl,Laz00prb,Gar02prb,Rya01prl,Rob10s}, wedge-shaped layers~\cite{Zdr06prl,Zdr13prb}, and more complex geometries~\cite{Gir98prb,Sos06prl,Alm09prb,Pet99prl,Pet00jltp}. In particular, flat geometries are well suited for observing the variation of the critical temperature $T_c$ of S on the thickness of F, while wedged layers~\cite{Zdr06prl,Zdr13prb} have been used for investigations of the triplet spin-valve effect caused by a non-collinear alignment of the magnetization of F layers. Other experiments~\cite{Wan09prl,Wan10nat,Liu12pcs} have been carried out in the nanowire geometry where marked drops in the resistance $R(T)$ of F were observed when sweeping the temperature below $T_c$ of S. At this point we would like to put the nanowire geometry in the focus of our presentation and to mention two issues typical experiments in this geometry share: Firstly, so far there has been few work studying the spatial extent of the superconducting proximity effect in one and the same nanowire. In addition to this, various preparation techniques were usually used for the nanowire fabrication in different works. As a consequence of this, the microstructure of samples markedly varied from work to work, making a study of the superconducting proximity effect at different length scales difficult. Secondly, as nanowires are fragile objects whose degree of defects is particularly sensitive to the preparation of macroscopic leads needed for electrical resistance measurements, the impact of contacts on the nanowire's conducting properties is hard to control. This is why, in the present work we used electrochemical deposition as a single fabrication technique for the preparation of metallic and ferromagnetic nanowires of well-defined microstructural properties. This allowed us to comparatively study the superconducting proximity effect in the different sections of one and the same individual nanowires and to compare the proximity effects at S/N and S/F interfaces. In addition to this, to keep the impact of contacts on the nanowire's electrical transport properties as identical as possible, the same direct-writing techniques by focused particle beams were used for contacting all the samples.

\begin{figure}
\centering
    \includegraphics[width=0.45\textwidth]{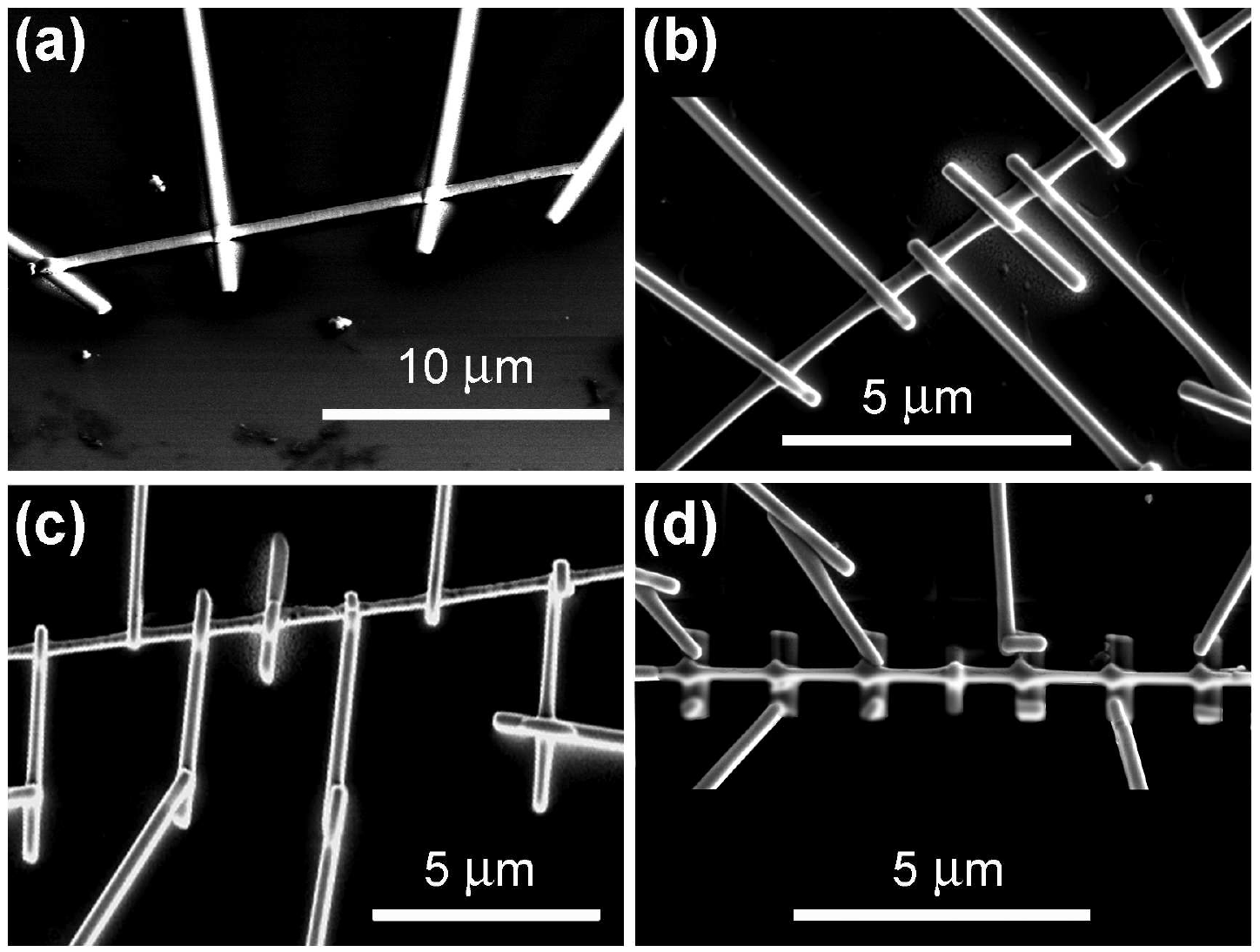}
    \vspace{0.5cm}\hspace{0.3cm}
    \includegraphics[width=0.45\textwidth]{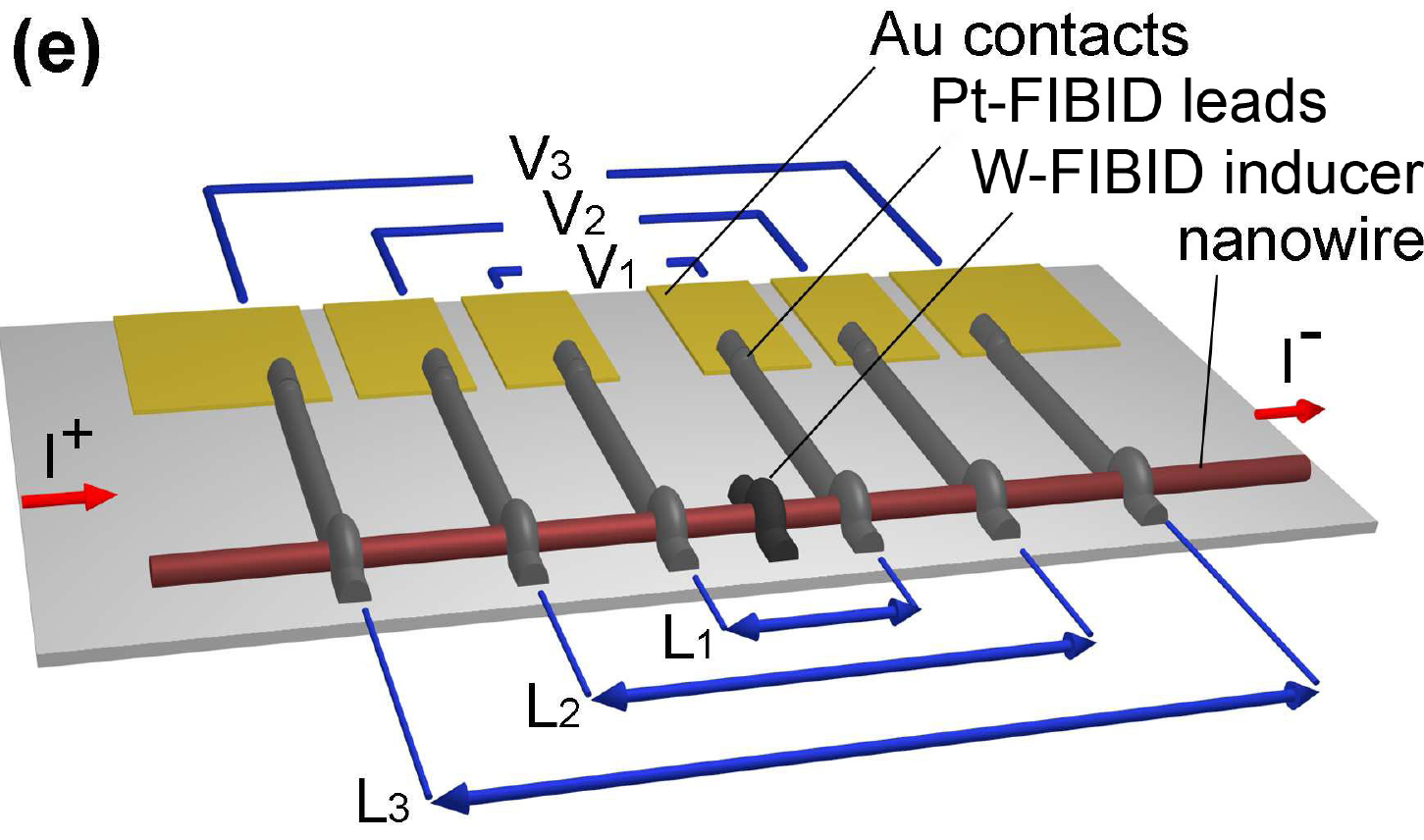}
    \caption[]
    {SEM images of the fabricated samples: Cu-NW1~(a), Cu-NW2~(b), Co-NW~(c), and Co-FEBID~(d). In the 4-probe geometry (a), all the electrodes are made from superconducting W-FIBID. The respective electrical scheme (e) of the 8-probe geometry realized in (b-d). In the 8-probe geometry, only the inner floating electrode is made of superconducting W-FIBID.}
   \label{fSEM}
\end{figure}

The geometry used for electrical resistance measurements in our experiment [see figure~\ref{fSEM}(e)] was an advanced nanowire geometry in which, along with the outer current contacts, up to three pairs of voltage leads have been attached to the nanowires in addition to the inner W-based superconducting floating electrode (inducer). This geometry is advantageous as compared to other geometries. Namely, (i)~it allows for measuring the nanowires' electrical resistance at different distances from the inducer, thereby allowing for the evaluation of the spatial extent of the superconducting order parameter in the nanowire. (ii)~The proposed geometry allowed us to perform measurements on different sections of the same nanowires that eliminated the problem of reproducibility of the microstructural sample properties. (iii)~This arrangement is very sensitive to the proximity effect due to the elongated nanowire geometry with a large aspect ratio. (iv)~The superconducting proximity effect can be studied on nanowires with a wide range of geometrical, microstructural, and compositional properties, as the nanowire fabrication techniques are well established \cite{Nee11rsj,Nee12rsj,Mut12bjn,Enc06apa,Toi01am,Toi12bjn}.
\begin{table*}
   \centering
   \begin{tabular}{l*{9}{c}}
	Object     & Structure      & Cross-section, nm        & Metal con-    & O,       &  C,      & Ga,      & Geometry\\
               &                & $\oslash$ or $w\times d$ & tent, at.~$\%$& at.~$\%$ & at.~$\%$ & at.~$\%$ &         \\
\hline
    Cu-NW1     & single-crystal & $\oslash$ 380            & $\times$ & $\times$ & $\times$ & $\times$ & figure~\ref{fSEM}(a)\\
    Cu-NW2     & single-crystal & $\oslash$ 275            & $\times$ & $\times$ & $\times$ & $\times$ & figure~\ref{fSEM}(b)\\
	Co-NW      & polycrystalline& $\oslash$ 280            & $\times$ & $\times$ & $\times$ & $\times$ & figure~\ref{fSEM}(c)\\
	Co-FEBID   & nanogranular   & $155\times255$           &    71    & 14       & 15       & $\times$ & figure~\ref{fSEM}(d)\\
\hline
    W-FIBID    & amorphous      & $150\times200$           &   47    & 8        & 30       & 16    &\\	
    Pt-FIBID   & amorphous      & $150\times180$           &   32    & 5        & 53       & 10    &\\
	
   \end{tabular}
   \caption{The geometrical and compositional parameters of the investigated nanowires and supplementary electrodes. $w$: width; $d$: thickness.}
   \label{tEDX}
\end{table*}

Here, we report the results of our experimental study of the superconducting proximity effect by electrical transport measurements, in three different types of metals in the nanowire geometry. The samples are two crystalline Cu, one Co nanowire and one Co nanogranular structure. These materials have been chosen for the following reasons: Cobalt is a strong ferromagnet for which investigations of the proximity effect are challenging due to a strong $h_{ex}$. The high-quality Cu nanowires were used as a reference diamagnetic system with the purpose of comparing the proximity effects in both metals. Our key observation is that the proximity effect in the crystalline cobalt nanowire is long-ranged, while this effect was not observed in the nanogranular Co structure, due to the dominating Cooper pair scattering caused by its intrinsic microstructure. The proximity-induced relative resistance drops in the polycrystalline Co nanowire have been revealed to be of the same order of magnitude as those in the Cu nanowire. Moreover, whereas these drops vanish with increasing magnetic field in the case of Cu, proximity-induced superconductivity in the polycrystalline Co remains unsusceptible to magnetic fields up to $11$~T (critical field of the inducer electrode at $T_c/2$), attesting to the spin-triplet nature of the observed long-ranged effect. In addition to that, an Arrhenius analysis of the proximity-induced relative resistance drops has revealed two different thermally-activated processes in the Cu nanowires, which we attribute to the contributions of intact and contact-damaged regions to the electrical resistance of the investigated samples.

The paper is organized as follows. The procedures of nanowire synthesis and contact preparation are described in section~\ref{sExperimental}. The results of transport measurements are reported in section~\ref{sResults}. A discussion of the data follows in section~\ref{sDiscussion}. The proximity lengths are quantified in section~\ref{sXi}. Conclusions round up the presentation of our results in section~\ref{sConclusion}.

\section{Experimental}\label{sExperimental}

\subsection{Nanowire synthesis}
The samples investigated in this work are two single-crystal Cu nanowires prepared by electrochemical deposition~(ECD) and two Co nanowires, one polycrystal prepared by ECD and one nanogranular prepared by focused electron beam induced deposition~(FEBID). Throughout the text the samples will be referred to as Cu-NW1, Cu-NW2, Co-NW, and Co-FEBID, respectively. Their structural and compositional parameters are compiled in table~\ref{tEDX}.

The cylindrical Cu and Co nanowires were grown in heavy-ion-track etched polycarbonate templates~\cite{Nee12rsj} whose thickness of $60$~$\mu$m determined the  maximal nanowire length. The ion tracks in the membranes were further etched up to $275-380$~nm in diameter and the nanowires were grown within the thus obtained pores. For further details of the employed ECD processes, we refer to~\cite{Nee11rsj,Nee12rsj}. The microstructure of the nanowires embedded in the membrane was investigated by means of X-ray diffraction. The acquired data confirmed the single-crystallinity of the Cu nanowires and polycrystallinity of the Co nanowire. Cu has been found to grow $(110)$-oriented along the nanowire axis, while $(100)$, $(110)$, and $(103)$ preferential orientations have been observed in the hexagonal close-packed (hcp) phase of Co. In the X-ray diffractogram for Co-NW, a small peak stemming from Co$_3$O$_4$ was also detected. After the X-ray measurements the nanowires were released from the templates by dissolving the membranes in dichloromethane.

For contacting individual nanowires, we used $p$-doped Si(100)/SiO$_2$ substrates with Cr/Au contacts of $3/100$~nm thickness prepared by photolithography in conjunction with lift-off. A small drop of the dichloromethane solution with nanowires was placed onto the substrate and, once dried-up, the substrate was mounted into a scanning electron microscope~(SEM). By SEM scanning along the nanowire axis it was possible to observe individual crystallite grains with a size of $400-700$~nm in Co-NW. We assume that Co$_3$O$_4$ was formed at the grain boundaries.

\subsection{Preparation of electrodes}
The SEM used in this work was a high-resolution dual-beam instrument~(FEI, Nova NanoLab 600) equipped with a multi-channel gas injection system for focused ion beam induced deposition (FIBID)~\cite{Utk08jvstb} and focused electron beam induced deposition~(FEBID)~\cite{Utk08jvstb,Hut12bjn}. These techniques allow for mask-less writing of predefined patterns with resolution in the nanometer range. In this way, we were able to make contacts of suitable conductance to the nanowires.

FIBID of Pt was used for the preparation of voltage leads. The precursor gas was $\mathrm{(CH_3)_3Pt(CpCH_3)}$, the beam parameters were $30$~keV/$10$~pA, the pitch was $30$~nm, the dwell time was $200$~ns, and the process pressure was $1.21 \times 10^{-5}$~mbar. A metal-insulator transition is known to occur in Pt-FIBID structures at liquid-helium temperatures once the deposit thickness is reduced below $50$~nm~\cite{Fer09prb}. Accordingly, Pt-FIBID leads with a thickness of $130-230$~nm and a width of $100-200$~nm were used in this work. Such dimensions ensured that the deposit's electrical conductivity is in the metallic regime.

FIBID of W was used for the preparation of superconducting inducer floating electrodes. The W-FIBID deposit is an amorphous W-based superconductor~\cite{Sad04apl}, with contributions of C and Ga, see table~\ref{tEDX} for the elemental composition. It has a critical temperature $T_c$ of $4.6 - 5.2$~K, depending on the deposition conditions (for comparison, $T_c \approx 0.012$~K for bulk W). In our work, for all the samples the precursor gas was $\mathrm{W(CO)_6}$, the beam parameters were $30$~keV/$10$~pA, the pitch was $18$~nm, the dwell time was $200$~ns, and the process pressure was $1.83 \times 10^{-5}$~mbar. As the gallium ion beam is known to cause amorphization, implantation and vacancy generation in the near-surface area of the exposed region~\cite{Dob12njp}, imaging of the nanowires and the electrodes with the ion beam was minimized. At this point it should be noted that the gallium itself is not responsible for superconductivity in W-FIBID electrodes, since it is also present in Pt-FIBID electrodes which are not superconducting.
\begin{table*}
   \centering
   \begin{tabular}{l*{14}{c}r}
Sample &  Current,& L$_1$, & L$_2$,   & L$_3$,   & $R_{1_{6\mathrm{K}}}$, & $R_{2_{6\mathrm{K}}}$, & $R_{3_{6\mathrm{K}}}$, & $\Delta R_1$, & $\Delta R_2$, & $\Delta R_3$, & $\xi_1$, & $\xi_2$, & $\xi_3$,\\
       &  $\mu$A  & $\mu$m & $\mu$m   & $\mu$m   & $\Omega$   & $\Omega$  & $\Omega$      & $\%$    & $\%$      & $\%$          & $\mu$m   & $\mu$m   & $\mu$m \\
\hline
Cu-NW1 &  0.2     & 7.5    & $\times$ & $\times$ & 0.33       & $\times$  &     $\times$  & 45      & $\times$  & $\times$      & 1.6     & $\times$  & $\times$ \\
Cu-NW2 &  1       & 2.2    & 4.6      & 9        & 0.23       & 0.63      &      1.17     & 33      & 13        & 5             & 0.38    & 0.25      & 0.2\\
Co-NW  &  0.1     & 3.8    & 7.2      & 12       & 3325       & 3575      &      5850     & 22      & 28        & 5             & 0.41    & 0.9       & 0.32\\
Co-FEBID& 0.5     & 2.1    & 4.8      & 7.5      & 42         & 160       &      265      & 5       & 5         & 5             & $\times$ & $\times$ & $\times$ \\
   \end{tabular}
   \caption{The electrical parameters and the deduced at $2.5$~K proximity lengths for all the samples.}
   \label{tALL}
\end{table*}

FEBID of Co was used for the deposition of the Co-FEBID granular nanowire structure. The precursor was $\mathrm{Co_2(CO)_8}$, the beam parameters were $3$~keV/$90$~pA, the pitch was $5$~nm, the dwell time was $1~\mu$s, and the process pressure was $1.3 \times 10^{-5}$~mbar. Before the deposition, the chamber was evacuated down to $7\times10^{-6}$~mbar. Care has been taken to avoid spontaneous dissociation of the precursor gas molecules and autocatalytic deposition of Co on the SiO$_2$ surface~\cite{Mut12bjn}. For this, a Si/SiO$_2$/Si$_{3}$N$_4$ substrate was used instead of a Si/SiO$_2$ substrate. We decided to prepare a nanowire-shaped deposit supplemented with six pairs of additional transverse sidebranches. The entire specimen was fabricated in one single deposition process. This ensured that the sidebranches acted as contact pads for the Pt-FIBID voltage leads, thereby preventing irradiation of the main deposit by the ion beam.

The material composition in all the deposited structures was controlled by using energy-dispersive x-ray (EDX) spectroscopy in the same SEM after the deposition, without exposure of the samples to air. The EDX data are summarized in table~\ref{tEDX}. SEM images of the samples thus fabricated are shown in figure~\ref{fSEM}(a-d).

\subsection{Transport measurements}
Transport measurements were made in a helium-flow cryostat equipped with a $12$~T superconducting solenoid. The electrical resistance was measured as a function of temperature in the standard 4-probe geometry (Cu-NW1) and in the 8-probe geometry (all other samples), as is shown in figure~\ref{fSEM}(e). In the 4-probe geometry, all four electrodes were from superconducting W-FIBID. In the 8-probe geometry, three pairs of Pt-FIBID voltage leads were attached at different distances $L_1$, $L_2$, and $L_3$ (see table~\ref{tALL} for the numbers) beside the superconducting W-FIBID inner inducer electrode. In what follows, the voltage drops between the inner, middle, and outer pair of leads will be denoted as $V_1$, $V_2$, and $V_3$, and the respective resistances as $R_1$, $R_2$, and $R_3$.

The major portion of electrical resistance measurements was taken in the dc current mode. The dc current was supplied by a Keithley~2636A source-meter and the dc voltage was measured with an Agilent~34420A nanovoltmeter. Selected measurements were repeated with an ac current sourced from a lock-in amplifier which also served as an ac voltmeter. The ac data were essentially the same as those measured with the dc current.

\section{Results}\label{sResults}

\subsection{Cu nanowires}
\begin{figure}
\centering
    \includegraphics[width=0.45\textwidth]{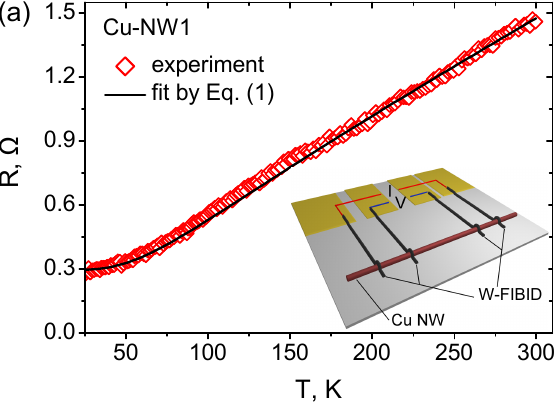}
    \vspace{0.5cm}\hspace{0.3cm}
    \includegraphics[width=0.45\textwidth]{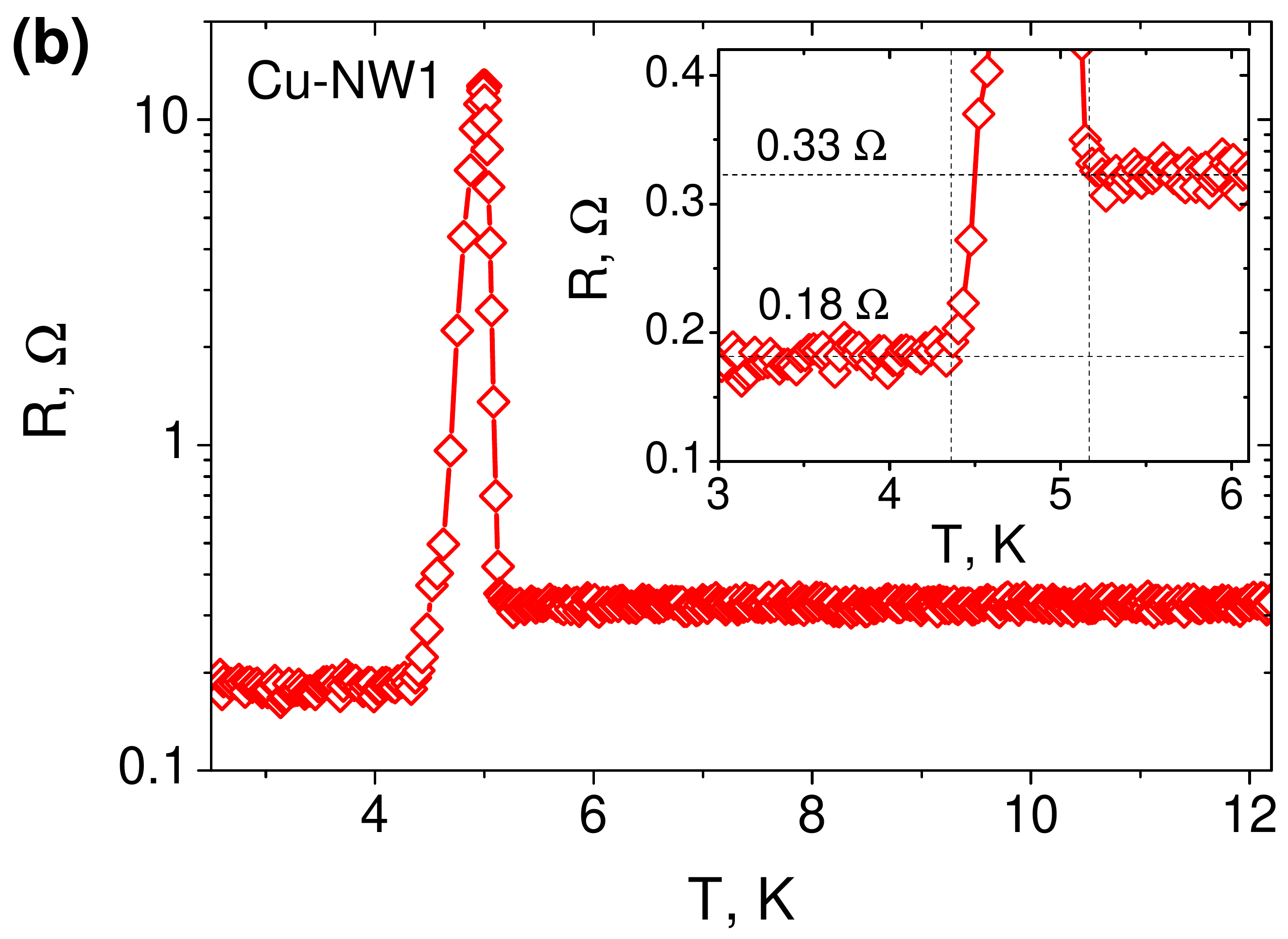}
    \caption[]
    {(a)~The temperature dependence of the resistance $R(T)$ for Cu-NW1. The straight line is a fit to the Bloch-Gr\"uneisen law by equation~\eqref{eBG} with $n = 5$ and a Debye temperature $\Theta_D = 343$~K~\cite{Kit04boo}. (b)~The $R(T)$ curve for the same nanowire close to $T_c$ of the superconducting electrodes. Below $5.2$~K, an anomalous resistance peak is observed. With further decreasing temperature, $R(T)$ levels off at $R \approx 0.55 R_{6\mathrm{K}}$.}
   \label{fCuNW1RvT}
\end{figure}

The temperature dependence of the electrical resistance $R(T)$ of Cu-NW1 is presented in figure~\ref{fCuNW1RvT}. Beside a pronounced peak in $R(T)$ in the vicinity of $5$~K, the dependence $R(T)$ demonstrates a typical metallic behavior: The curve has a practically linear section above $100$~K, a residual plateau below $25$~K, and a power-law crossover in between. The room-temperature resistivity of Cu-NW1 is $\rho_{295\mathrm{K}} = 2.2~\mu\Omega$cm, which is by $25\%$ higher than the literature value of $1.7$~$\mu\Omega$cm for bulk copper~\cite{Pow99jpc}. In figure~\ref{fCuNW1RvT}(b), the resistance peak at $T\approx5$~K is one-and-a-half order of magnitude larger than the resistance value at 6~K. Below $4.5$~K, the nanowire resistance is reduced by almost a factor of two as compared to $R_{6\mathrm{K}}$ and remains at this level down to 2.5~K being the lowest temperature achievable in our experiment.

\begin{figure}
    \centering
    \includegraphics[width=0.45\textwidth]{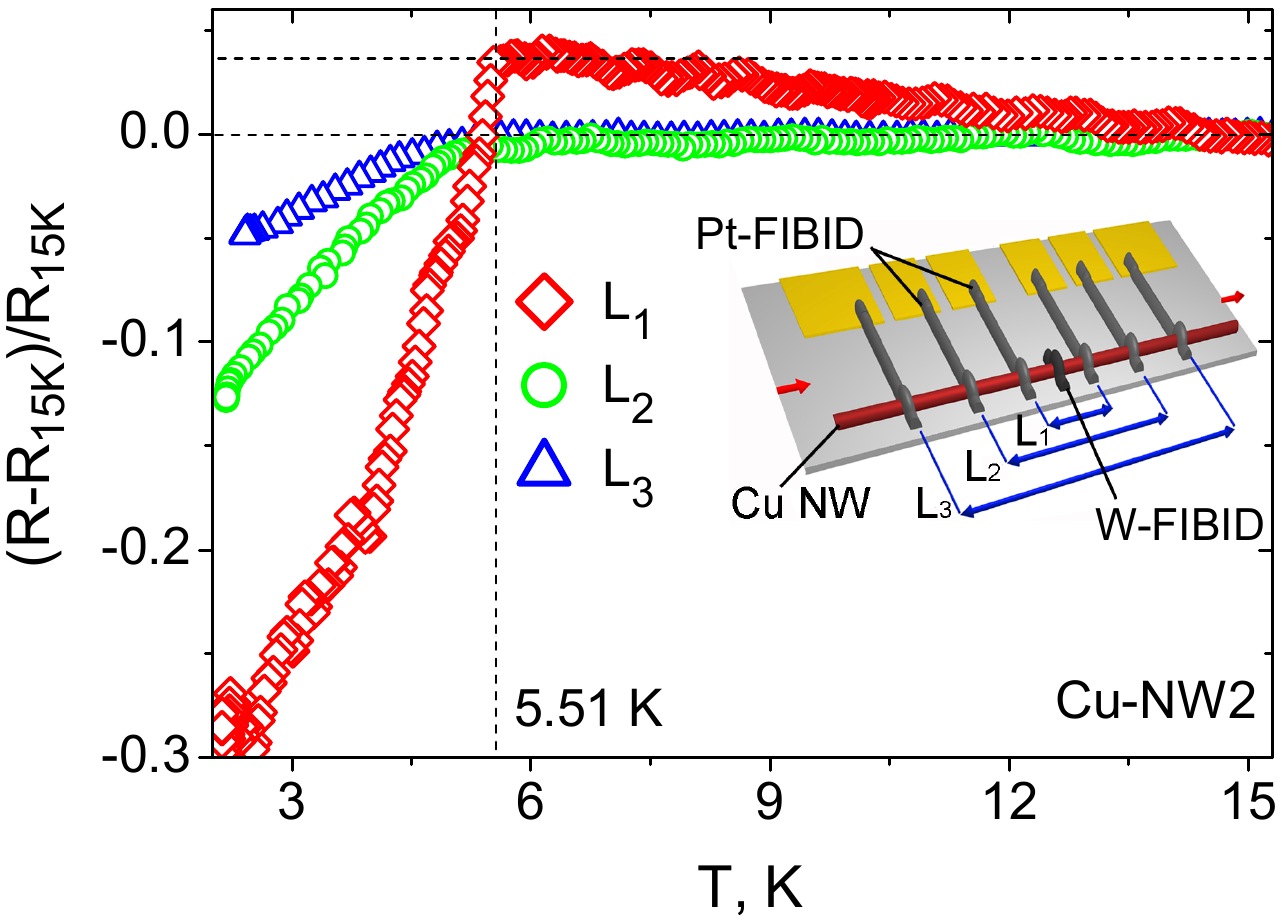}
    \caption[]
    {The temperature dependence of the relative resistance changes $(R-R_{15\mathrm{K}})/R_{15\mathrm{K}}$ for Cu-NW2. Inset: The layout of contacts.}
   \label{fCuNW2RvT}
\end{figure}

Now we turn to the presentation of $R(T)$ data for Cu-NW2. In the temperature range from $15$ to $295$~K, the behavior of $R(T)$ for Cu-NW2 is very similar to that of Cu-NW1 (not shown). By contrast, in the temperature range between $2$ and $15$~K the behavior of $R(T)$ for Cu-NW2 differs from that of Cu-NW1. These  distinctive features are presented in figure~\ref{fCuNW2RvT}, where $R_{15\mathrm{K}}$ is the resistance at $T=15$~K. Firstly, no resistance peak anomaly was observed in the vicinity of $5$~K for Cu-NW2. Secondly, the behavior of the $R(T)$ curves measured for the three nanowire sections differs substantially. With the reduction of $T$ from $15$ to $5.5$~K, a residual plateau is maintained for the outer and the middle voltage leads, while for the inner leads the resistance increases by 5\%. With further reducing the temperature from 5.5 to 2~K the resistance of all the sections decreases, with different temperature derivatives. Here, in contrast to the sudden resistance drop observed in Cu-NW1, the reduction of the resistance with decreasing temperature is much slower and the relative resistance changes are smaller ($5-35\%$ as compared to $45\%$ in the case of Cu-NW1). Finally, one may notice (refer to $L_1$ in figure~\ref{fCuNW2RvT}) two different slopes in $R(T)$ below and above 3.7~K.

\subsection{Co nanowire}
\begin{figure}
\centering
    \includegraphics[width=0.45\textwidth]{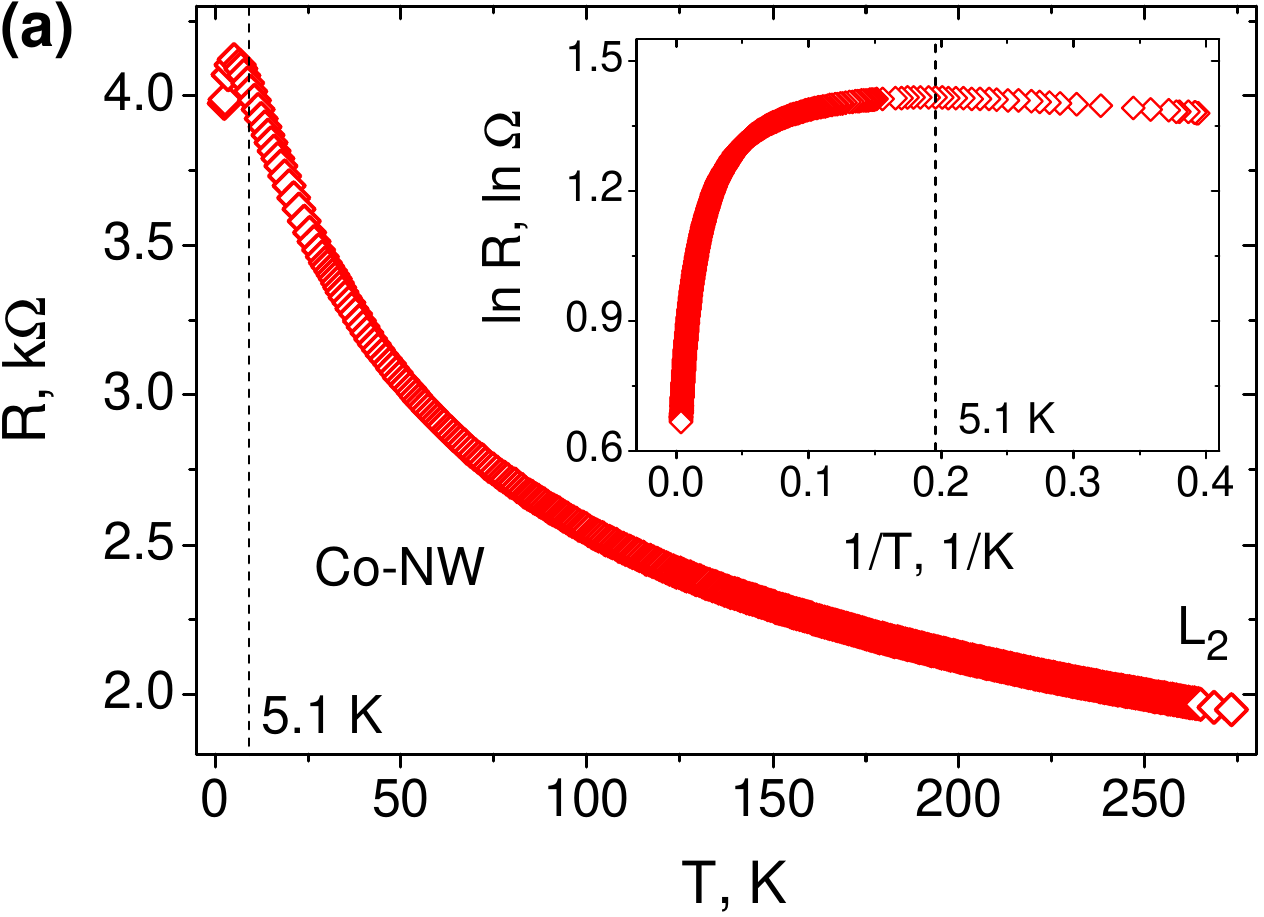}
    \vspace{0.5cm}\hspace{0.3cm}
    \includegraphics[width=0.45\textwidth]{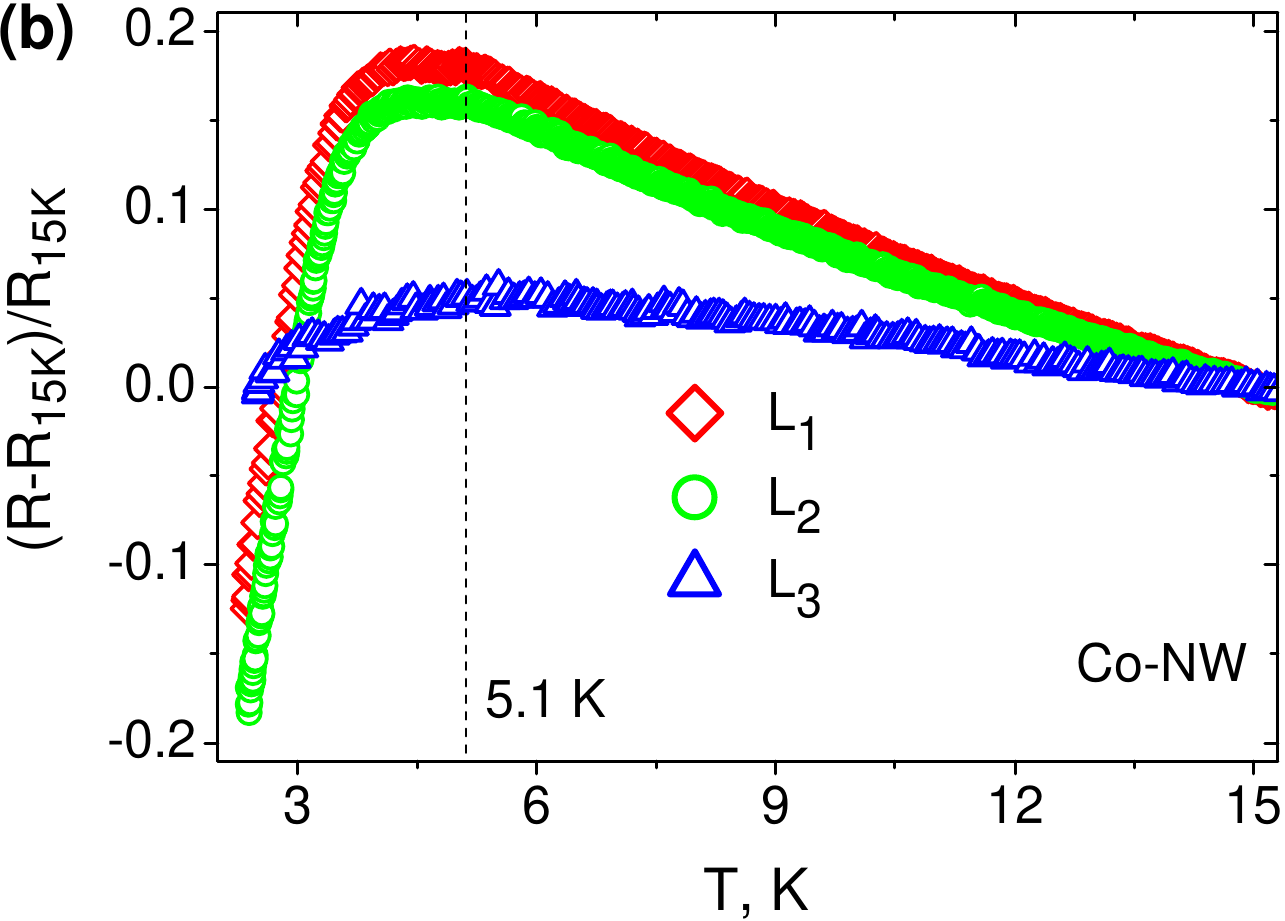}
    \caption[]
    {(a)~The temperature dependence of the resistance $R_2(T)$ for Co-NW. Inset: The same dependence in the $\ln R - 1/T$ coordinates. The vertical dashed lines mark the temperature where the curves start to deviate from  the localization behavior. (b)~Proximity effect-induced resistance drops for the three sections of Co-NW. The measurements were made in the same geometry as in the inset of figure~\ref{fCuNW2RvT}.}
   \label{fCoNWRvT}
\end{figure}
Figure~\ref{fCoNWRvT}(a) depicts the temperature dependence of the resistance of Co-NW for the $L_2$ section. The $R(T)$ dependence demonstrates a thermally activated behavior. The room-temperature resistivity of Co-NW $\rho_{295\mathrm{K}} = 1771~\mu\Omega$cm is two orders of magnitude larger than the literature value of about $5.8~\mu\Omega$cm for bulk Co~\cite{Kit04boo}. This high value of the resistivity is caused by the contribution of grain boundaries, which will be addressed in more detail in section~\ref{sDiscussion}.

The low-temperature resistance data for the different voltage probes are presented in figure~\ref{fCoNWRvT}(b). As follows from the figure, the superconducting proximity effect prevails over the localization behavior below $5.1$~K, as the curves start to deviate from the thermally activated behavior. With further reduction of temperature a rapid drop of the resistance follows for the inner and the middle voltage probes. The maximal resistance drop of about $28\%$ with respect to the normal resistance state was observed for the $L_2$ section.

\subsection{Co-FEBID structure}
\begin{figure}
\centering
    \includegraphics[width=0.45\textwidth]{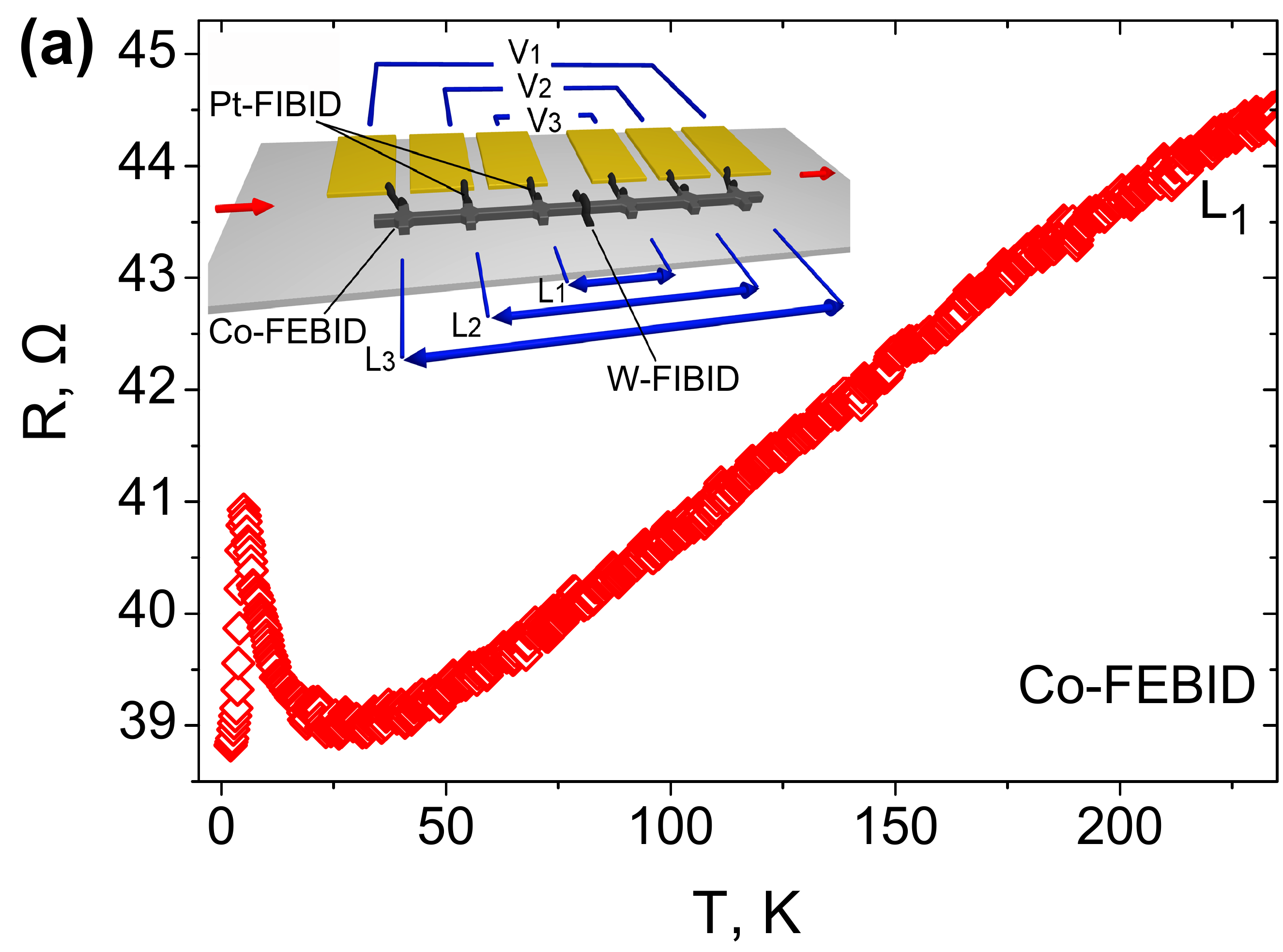}
    \vspace{0.5cm}\hspace{0.3cm}
    \includegraphics[width=0.45\textwidth]{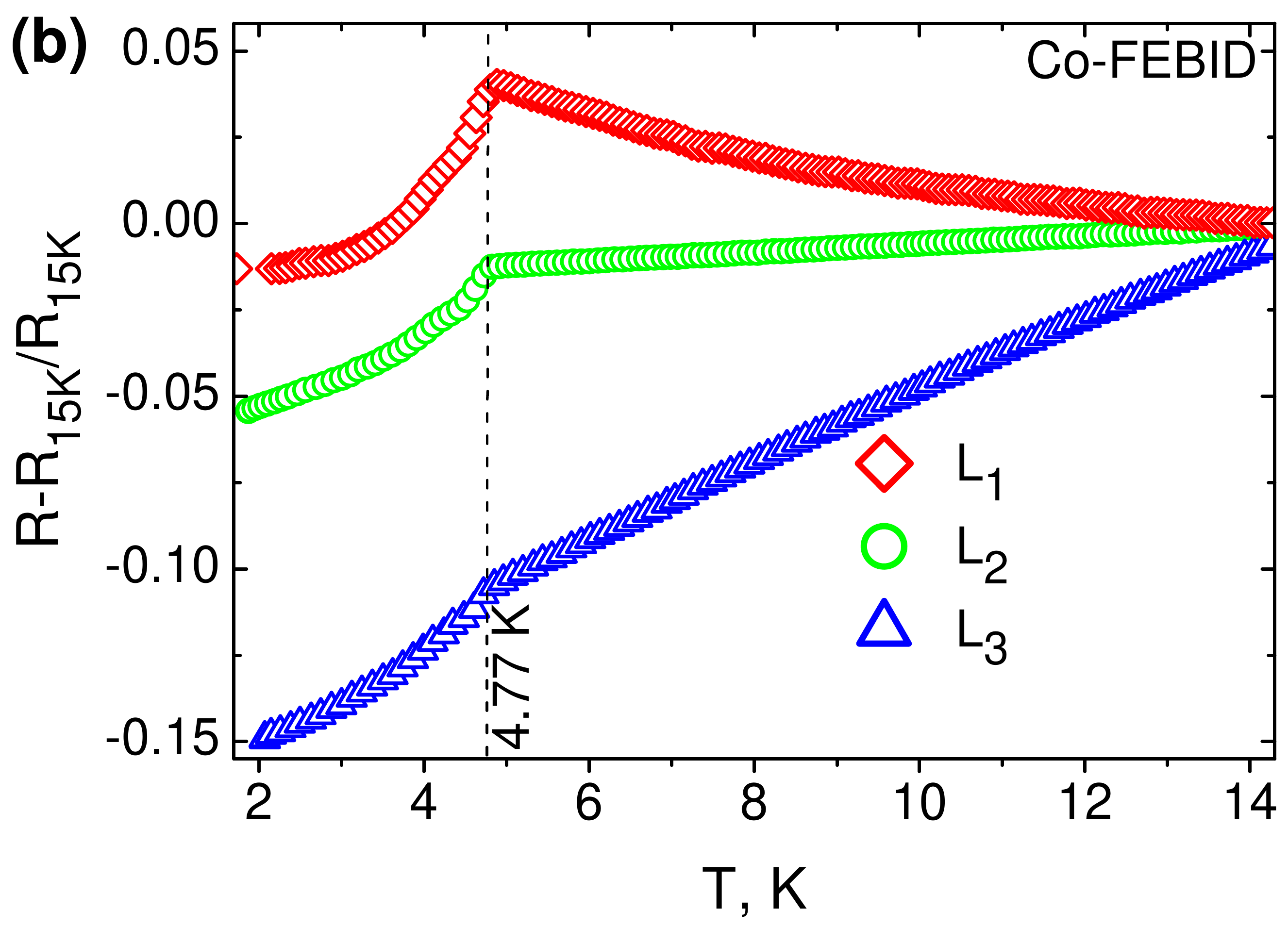}
    \caption[]
    {(a)~The temperature dependence of the resistance $R_1(T)$ for Co-FEBID. Inset: The layout of contacts. (b)~The relative resistance changes for the three sections of the Co-FEBID structure. The observed resistance drops are less pronounced as compared to Co-NW.}
   \label{fCoEBID}
\end{figure}
The temperature dependence of the resistance $R_1(T)$ for the Co-FEBID structure is shown in figure~\ref{fCoEBID}(a). The room-temperature resistivity of Co-FEBID is $\rho_{295\mathrm{K}} = 84~\mu\Omega$cm, i.e. approximately $15$~times larger than that of the reference bulk value~\cite{Kit04boo}. The cooling curve has a virtually linear, metallic-like section between 295~K and 30~K and demonstrates a tendency to localization at lower temperatures, followed by a resistance drop at $T \approx 4.77$~K. Figure~\ref{fCoEBID}(b) displays the low-temperature resistance data for the three pairs of potential probes. The temperature location of the resistance drops is by about $0.4$~K lower than those for Cu-NW2 and Co-NW. In addition to that, above $4.77$~K, the $R(T)$ curves demonstrate a qualitatively different behavior for the different sections. Namely, for the inner section $R(T)$ increases with decreasing temperature, while for the middle and outer sections $R(T)$ decreases with decreasing temperature, with different temperature derivatives. The relative resistance drops for $V_1$, $V_2$ and $V_3$ are only $\approx5\%$ with respect to $R_{4.77~\mathrm{K}}$.

\section{Discussion}\label{sDiscussion}

\subsection{Cu nanowires}
In this subsection, we first focus on the overall behavior of the cooling curves, next discuss the resistance peak anomaly, and finally analyze the observed resistance drops in the vicinity of 5~K in more detail.

The overall shape of the $R(T)$ curve for Cu-NW1 can be rather well fitted to the Bloch-Gr\"uneisen formula~\cite{Zim60boo,Bid06prb}
\begin{equation}
    \label{eBG}
    R(T) = R_0 + K(T/\Theta_D)^n \int_{0}^{\Theta_D/T}dx\frac{x^n}{(e^x-1)(1-e^{-x})},
\end{equation}
where $R_0$ is the residual resistance, $\Theta_D$ is the Debye temperature ($343$~K for Cu~\cite{Kit04boo}), and K is the only fitting constant. In equation~\eqref{eBG}, $n$ is an integer determining the power law which in turn depends on the prevailing scattering mechanism in the sample. The resulting fit is shown in figure~\ref{fCuNW1RvT}(a). The fitting parameter $K$ has been chosen such that the best possible coincidence with the experimental curve is achieved for $R_{25\mathrm{K}}$ and $R_{295\mathrm{K}}$. The $R(T)$ curve in figure~\ref{fCuNW1RvT}(a) has been fitted by equation~\eqref{eBG} with $n=5$ which implies that the resistance is due to scattering of electrons by phonons, as expected for nonmagnetic metals~\cite{Zim60boo,Bid06prb}. Taking into account the close-to-bulk residual resistance of Cu-NW1 we conclude that this nanowire represents a high-quality reference sample. Now we proceed to the discussion of the resistance peak anomaly.

An anomalous resistance peak in the vicinity of $T_c$ of S, similar to that reported in this work, has been observed in a number of experiments~\cite{Wan10nat,Aru99prb,Par95prl,Str96cjp}. For its explanation, several theoretical models~\cite{Lan97prb} have already been proposed. It is widely appreciated that the resistance anomaly and the superconducting proximity effect are two neighboring phenomena. Specifically, the anomalous resistance peak usually appears in the vicinity of $T_c$ and, with further decreasing temperature, a proximity effect-induced resistance drop takes place. Besides, for both the resistance peak and drop, the existence of an S/N interface is the necessary condition. As a generalization of the literature data~\cite{Wan10nat,Wan09prl,Cou95prb}, one can state that:
 \begin{enumerate}
   \item The appearance of the resistance peak is not related to the magnetic ordering of N, as the effect has been observed for superconductors in contact with ferromagnetic Co~\cite{Wan10nat} and Ni~\cite{Wan10nat}, as well as diamagnetic Au, Cu (this work), and Ag.
   \item The magnitude of the resistance peak and its form can show large variability even if the S-counterpart materials are the same.
   \item The effect depends on the nanowires length and the sample-contact interface(s).
   \item The effect is observed regardless of whether S is a part of the electric circuit or it is a floating electrode.
 \end{enumerate}
To explain the resistance peak in our data, we refer to the model suggested in~\cite{Lan97prb} which has already been confirmed experimentally~\cite{Aru99prb}. The main idea of that model is the following: It has been proposed~\cite{Aru99prb} that an increase of the resistance above the normal-state value at the top of the superconducting transition is due to a deformed N/S boundary in the vicinity of the voltage probes. The deformed N/S boundary leads to the formation of a non-equilibrium region inside S. The formation of the latter is caused by the injection of quasiparticles from N characterized by a finite value of the electric field and a corresponding effective resistance. Although our situation differs from that in experiment~\cite{Aru99prb}, some parallels can still be drawn. Here, we used the FIBID technique for contacting Cu-NW1 with two superconducting W electrodes. The thus obtained system has two N/S boundaries with unpredictable shapes in the vicinity of the contact regions whose asymmetry must be the cause of the observed resistance peak.

\begin{figure}
\centering
    \includegraphics[width=0.4\textwidth]{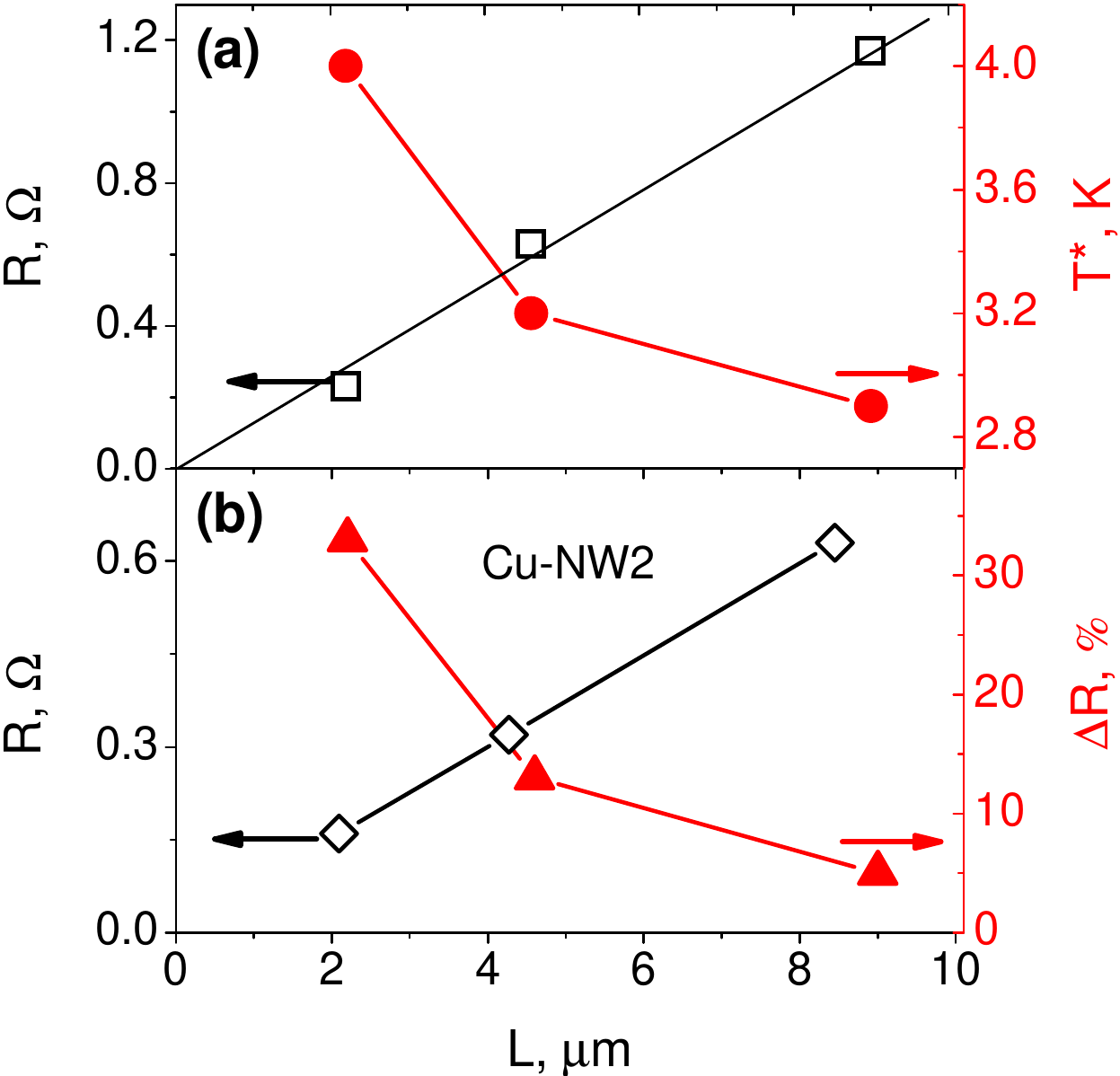}
    \caption[]
    {(a)~Left axis: Dependence of the Cu-NW2 nanowire resistance $R_{15\mathrm{K}}$ ($\Box$) on the measured section length $L$. The straight line is a linear fit with $\rho_{15\mathrm{K}} = 0.67~\mu\Omega$cm for an assumed ideal wire (cylindrical shape, ideal single crystal, homogenous current distribution, no damaged areas owing to the FIBID of contacts). Right axis: Variation of the crossover temperature $T^\ast$ versus $L$ ($\bullet$). (b)~Left axis: The nanowire resistance as a function of its length without contributions of the defect-rich contact areas ($\diamondsuit$), see text for details. Right axis: The relative proximity-induced resistance drops $\Delta R$ ($\blacktriangle$) versus the nanowire section length $L$. }
   \label{fCuNW2RvL}
\end{figure}

Now we turn to an analysis of the resistance drops in Cu-NW2. As the measurements were taken at small transport currents and we found the results to be independent of the transport current magnitude for $I\leq10~\mu$A, an Arrhenius' analysis can be applied to the temperature dependences of the resistances in figure~\ref{fCuNW2RvT} in order to check whether some activation mechanism can be identified at $T\lesssim5.5$~K. The Arrhenius analysis relies upon the assumption that the resistance of the sample is independent of the transport current and is given by Arrhenius' law
\begin{equation}
     \label{eArrh}
     R = R_0\exp\frac{-U}{T},
\end{equation}
\begin{figure}
    \centering
    \includegraphics[width=0.45\textwidth]{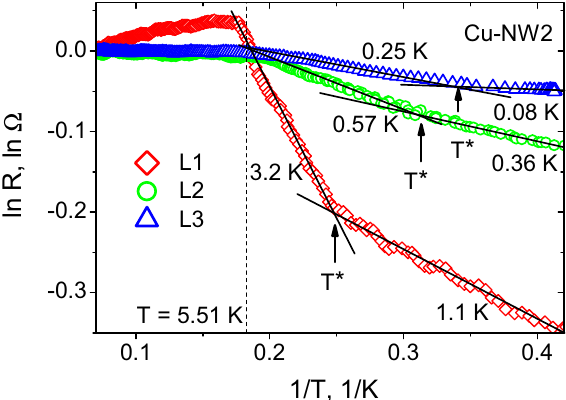}
    \caption[]
    {Arrhenius plots $\ln R (1/T)$ with the deduced activation energies~$U$ for the three nanowire sections of Cu-NW2. Two different activation processes can be identified for all the sections, with a crossover at $T^\ast(L)$.}
   \label{fCuNW2Arrh}
\end{figure}
where $R_0$ is a constant and $U$ is the activation energy of some process. Then, if one plots $\ln R$ vs $T^{-1}$ and this curve can be fitted by a straight line, which is the fingerprint of the thermo-activated character~\cite{Dob12njp,Sor07prb}, the slope of the linear part of the Arrhenius plot gives the activation energy $U$. The Arrhenius plots with $R_0 = R_{15\mathrm{K}}$ are shown in figure~\ref{fCuNW2Arrh}. The corresponding activation energies for the different nanowire sections are labeled close to the curves. Evidently, for each section \emph{two different} activation processes take place. The activation energy of the process dominating at close-to-critical temperatures is higher than that prevailing at far-subcritical temperatures. If we denote the temperature corresponding to the crossover between the two different activation processes as $T^\ast$, its value is $4$~K, $3.2$~K, and $2.9$~K for the sections $L_1$, $L_2$, and $L_3$, respectively~(figure~\ref{fCuNW2RvL}(a), right axis). We attribute the two thermo-activated processes to the contributions stemming from two different regions of the nanowire. Namely, one contribution originating from the nanowire region which has been damaged by FIBID and another from unaffected regions. As the temperature decreases below $T_c$, the intact part first comes to the low-resistance state. For this part the drop of $R(T)$ is most steep and this corresponds to a larger activation energy in figure~\ref{fCuNW2Arrh}. Turning to the damaged-part contribution, we note the the reduction of $R(T)$ is less steep which we attribute to an irregular defect distribution. This mechanism corresponds to a lower activation energy and is reflected in a long nonzero resistance ``tail`` down to the lowest achievable temperature. This can be explained by the fact that there is an incomplete superconducting phase coherence over the measured nanowire sections even at temperatures far below $T_c$ of the inducer electrode. This incomplete phase coherence is likely caused by thermally induced phase slippage in inhomogeneous regions caused by the FIBID process. The shift of the crossover temperature $T^\ast$ for the different sections may be explained by different ratios of the ion beam-damaged volume fraction to the total volume of the measured nanowire section, while the higher activation energy of the thermo-activated process corresponds to the lower degree of disorder. In this way, we have the \emph{two superimposed effects} of proximity-induced resistivity drops and nanopatterning-caused disorder in the ion beam-exposed regions.

\begin{figure}
\centering
    \includegraphics[width=0.35\textwidth]{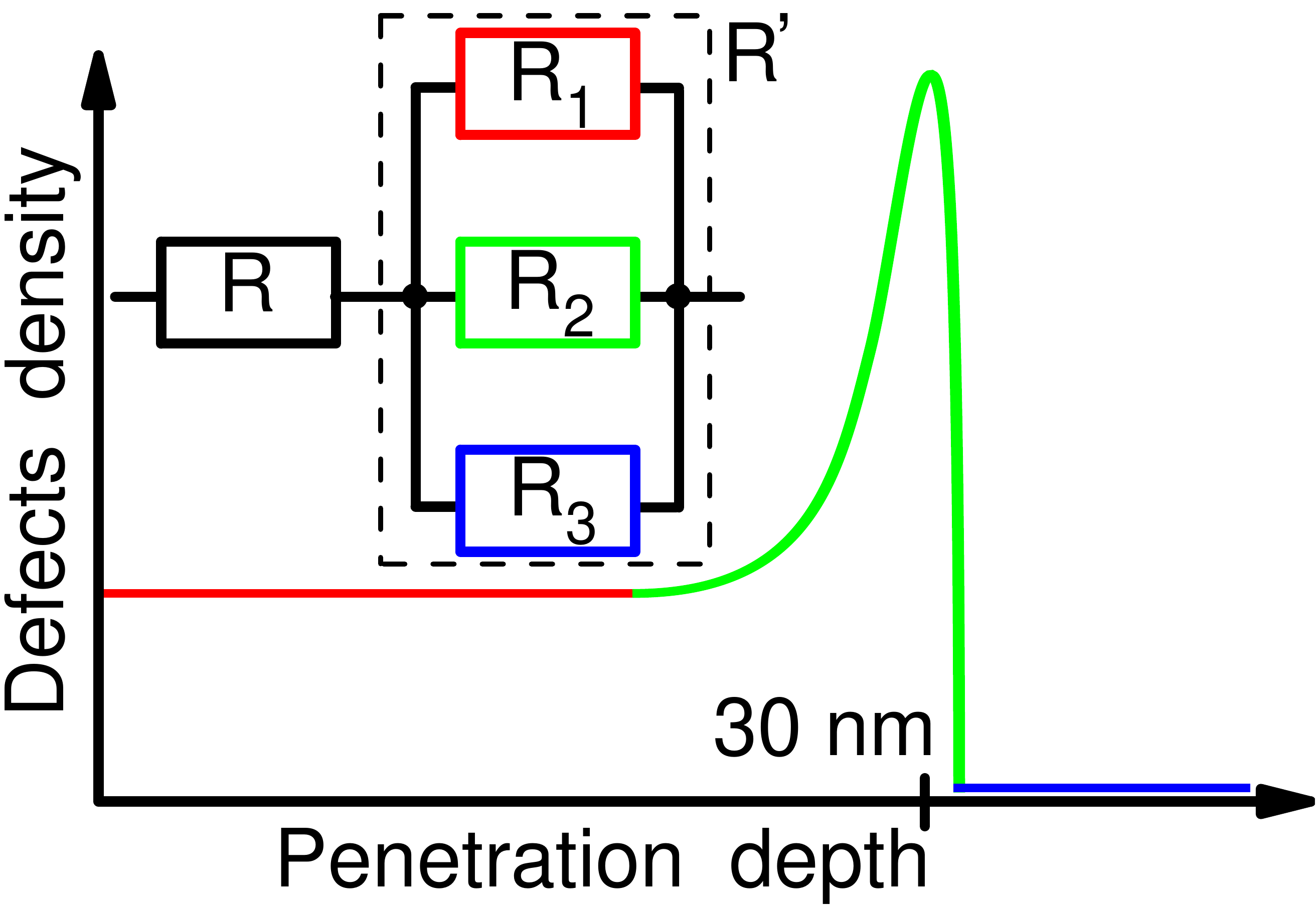}
    \caption[]
    {Sketch of the resistance model used for quantifying the changes in the nanowire resistivity due to the generation of defects during the FIBID processing. The distribution of defects behind the ion track follows the Bragg curve which has a plateau in the quasi-ballistic regime of the ion motion and a pronounced peak in the diffusive regime. With further increasing depth, the density of produced defects drops down to zero. This corresponds to the virtually undamaged region preserved after the dominant portion of ions has been stopped at the characteristic penetration depth which is estimated as $30$~nm in the present case.}
   \label{Rmodel01}
\end{figure}

We now consider the normal-state resistance changes in Cu-NW2 during the processing by FIBID in more detail. During the FIBID process some part of the nanowire underneath the W and Pt electrodes is irradiated by Ga ions. Using simulations by the Monte Carlo method~\cite{Jam08boo} we estimate that the ions penetration depth is $30$~nm for an ion beam energy of $30$~keV. This means that the conducting properties of a part of the nanowire with a maximal layer thickness of $30$~nm have been changed. To account for the resistance changes due to the FIBID processing, we use the model circuit shown in the inset of figure~\ref{Rmodel01}. In this circuit, the entire nanowire is modeled as a series of affected and unaffected parts. The unaffected part is denoted as resistor $R$ while the situation becomes more complicated for the part exposed to the ion beam. This is because of the strongly inhomogeneous distribution of the defects behind ions passing through the material. The distribution of defects is described by the Bragg curve~\cite{Sig06boo}, as is depicted in figure~\ref{Rmodel01}.
By using this distribution law for the nanowire irradiated regions one can distinguish three different types of defect degrees across the nanowire cross-section: (i) The defect density is virtually constant in the regions where the ballistic mode prevails in the motion of ions~(red part of the Bragg curve in figure~\ref{Rmodel01}). (ii) An enhanced density of defects occurs corresponding to the Bragg peak in the region where the ions are stopped~(green part). (iii) The undamaged internal region of the nanowire~(blue part). Each part of the Bragg curve corresponds to a nanowire region with resistance values $R_1$, $R_2$ and $R_3$, respectively, which all together define the effective resistance $R'$ of the damaged region. As the exact geometry of the three mentioned parts of the nanowire is hard to estimate, it was impossible to calculate the respective resistivities separately. However, we have succeeded in quantifying the resistivity eigenvalue of the intact nanowire and the resistivity of the regions under the Pt- and W-FIBID electrodes using $30$~nm as a rough estimate for the ions' penetration depth. In our calculations we assume that the nanowire has an ideal cylindrical shape and the current distribution is homogeneous. The central result of our analysis is that the resistivity values for the regions underneath the Pt and W electrodes at $6$~K are $\rho^* = 4.2~\mu\Omega$cm and $\rho^{**} = 6.2~\mu\Omega$cm, respectively, being an order of magnitude higher than the resistivity of the undamaged Cu nanowire $\rho = 0.45~\mu\Omega$cm. As a proof of our calculations on the basis of this simple model, we recall that the resistivity of Cu-NW1 unexposed to FIBID over the entire measured section has virtually the same value $\rho = 0.51~\mu\Omega$cm at $6$~K. This fact allows us to conclude that the conducting properties of Cu-NW1 and the undamaged regions of Cu-NW2 are very similar and the employed model reasonably describes the changes in the conducting properties of the nanowires during their processing by FIBID. Finally, having subtracted the calculated contributions of the ion-beam damaged regions, the nanowire resistance versus its length is plotted in figure~\ref{fCuNW2RvL}(b)~(left axis). One can notice a factor of two reduction of the wire resistance as compared to the as-measured values presented in figure~\ref{fCuNW2RvL}(a)~(left axis).

Summarizing up to this point, our analysis of the resistance data for Cu-NW2 has shown that the relative defect concentration for the $L_1$ section is larger than that for $L_2$ and $L_3$. At the same time, the proximity-induced resistance drop for the $L_1$ section is most pronounced due to the shortest distance to the W inducer electrode. In this way, our model explains why the $L_1$ section has the highest $T^\ast$ and the largest activation energy at close-to-critical temperatures.

\subsection{Co nanowire}

\begin{figure}
\centering
    \includegraphics[width=0.45\textwidth]{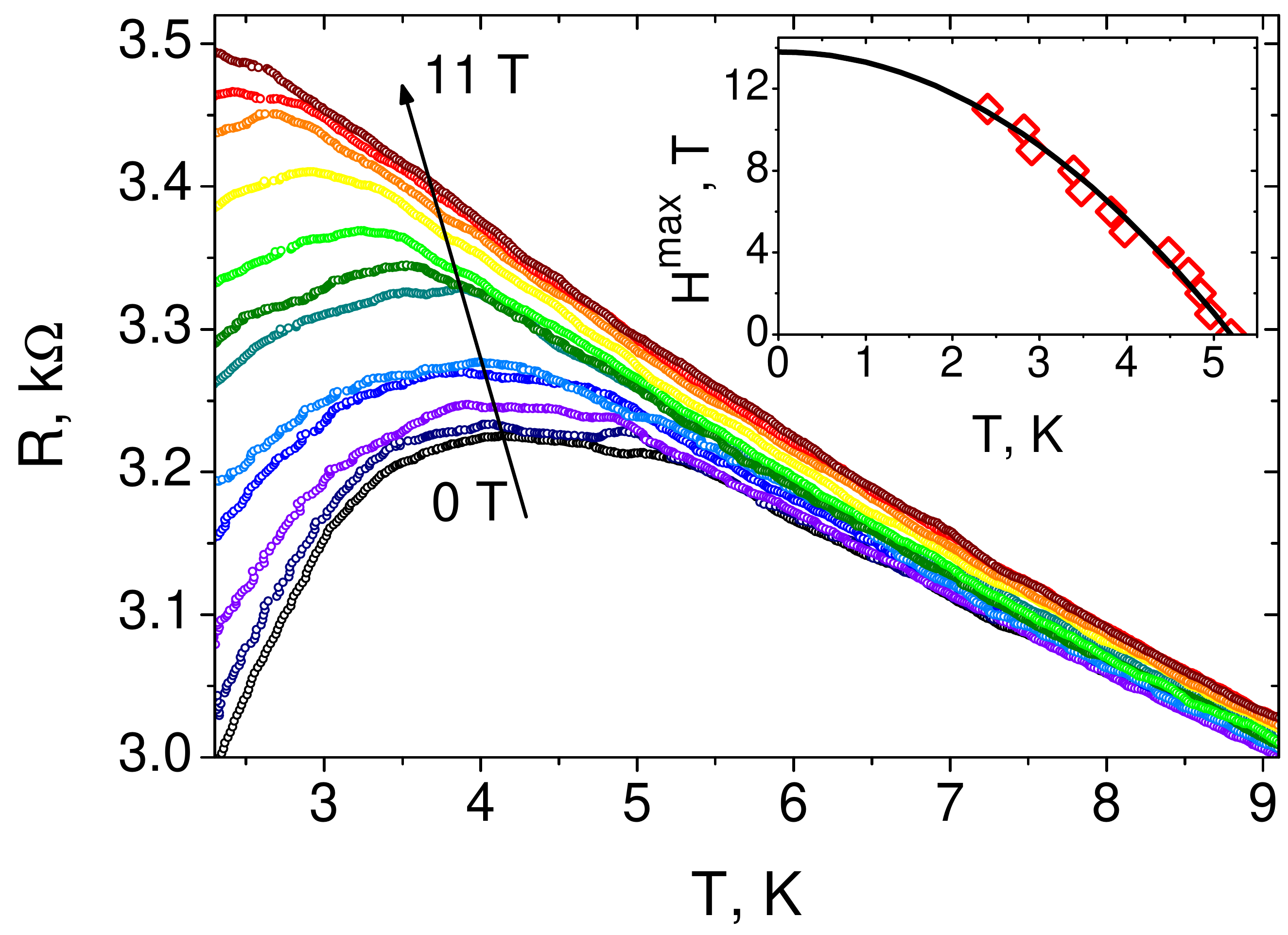}
    \caption[]
    {The temperature dependance of the resistance $R_1(T)$ for Co-NW for a set of applied magnetic fields from $0$ to $11$~T with a step of $1$~T along the black arrow. Inset: The dependence of the magnetic field $H^{\mathrm{max}}$ versus $T$ deduced from the resistance maxima at different fields in the $R(T)$ curves. The solid line is a fit to equation~\eqref{HvT} with $H_c(0)$ = $13.5$~T attesting to that it is the (upper) critical field of the superconducting inducer electrode.}
   \label{HcvT}
\end{figure}

In section~\ref{sResults} it was shown that the resistance drops for Co-NW in the vicinity of $5$~K are of the same order of magnitude as those for Cu-NW2. The magnitude of the $R(T)$ drops clearly indicate that the superconducting proximity effect in Co-NW is long-ranged. Further support of this conclusion can be obtained from the magnetic field dependence of the proximity effect. Namely, the ``classical'' proximity-induced superconductivity is already suppressed in fields far below the upper critical field of the inducing superconductor. By contrast, if the pairing is spin-triplet, proximity-induced superconductivity in the ferromagnet should survive as long as the critical field of the superconductor is not reached. For this reason, measurements of $R(T)$ of Co-NW in magnetic fields up to $11$~T have been performed. The field was aligned in the substrate plane at an angle of $67^\circ$ with respect to the nanowire axis. The oblique angle was a result of the accidental orientation of the nanowire with regard to the pre-formed contact pads. The measured $R(T)$ curves for the $L_1$ section are presented in the main panel of figure~\ref{HcvT}. One can clearly see that, with increasing magnetic field, the temperature of the $R(T)$ maximum shifts towards lower temperatures, but the drop itself is maintained up to 10~T. Besides, the superconducting proximity effect competes with the localization behavior in even higher magnetic fields. We find that the data points $H^{\mathrm{max}}(T)$, obtained by plotting the field values versus the temperature at which the $R(T)$ curves have their respective maxima, nicely follows the empirical law~\cite{Buc04boo}
\begin{equation}
   \label{HvT}
   H^{\mathrm{max}}(T) = H_c(0)[1-(T/T_c)^2],
\end{equation}
where $H_c(0)$ = $13.5$~T is the upper critical field and $T_c = 5.2$~K is the superconducting transition of the inducer. The resulting fit is shown in the inset of figure~\ref{HcvT} by the solid line.
For other sections, the $H^{\mathrm{max}}(T)$ curves can also be fitted well by equation~\eqref{HvT} with the same $H_c(0)$ and $T_c$ (not shown). Apparently, the disappearance of proximity induced superconductivity is not due to pair-breaking effects in Co-NW, but rather due to the breakdown of superconductivity in the W inducer itself. Theqrefore, we arrive at the conclusion that the observed effect is inspired by spin-triplet pairing unsusceptible to the magnetic field.

We now direct our attention to those microstructural properties of Co-NW which made it possible to observe the long-ranged spin-triplet proximity effect. As it is known from the literature~\cite{Hen01epjb} the crystallographic texture of Co nanowires grown by ECD depends on the nanowire diameter. If the nanowire diameter is smaller than a critical diameter $\oslash_c\approx50$~nm, a single-crystal microstructure ensues, while a polycrystalline microstructure is observed for $\oslash > \oslash_c$~\cite{Hen01epjb}. In both cases the hcp structure is realized. In addition to that, it is worth noting that for nanowires with $\oslash < \oslash_c$ the magnetization is mostly oriented longitudinally and a single-domain state is the ground state~\cite{Hen01epjb}. By contrast, for larger diameters $\oslash > \oslash_c$, a complex multidomain state is energetically favorable. In the multidomain state, the domain magnetization is oriented transverse to the nanowire axis. Since our nanowire is thick enough ($\oslash = 280$~nm) in terms of the above criterion, it has a polycrystalline microstructure, as confirmed by both x-ray diffraction measurements and a direct SEM inspection, and a multidomain state has to be assumed for the ground state of Co-NW. Naturally, domain boundaries are sources of a magnetization inhomogeneity, i.e. the necessary condition for the formation of the spin-triplet pairing in the ferromagnet. In addition to that, the employed contacting procedure by FIBID implies producing contact-damaged regions underneath the leads which we believe amplify the magnetization inhomogeneity even further. Besides, as a peak stemming from Co$_3$O$_4$ was detected in the x-ray data, we assume that cobalt oxide is located at the boundaries of individual crystallites. This assumption is in line with the electrical resistance measurements where a clear tendency to localization behavior has been observed. This corresponds to a thermally assisted electron tunneling regime between neighboring grains separated by a cobalt oxide layer. It is this tunneling regime which causes the observed high-resistance state of Co-NW.

\subsection{Co-FEBID structure}

Analogously to the previous subsection, here we analyze the microstructural properties of the Co-FEBID nanowire structure. At this, the driving question is why proximity effect-induced superconductivity does not become apparent in this sample. To answer this question, we first note that in the case of Co-FEBID we deal not with a homogeneous ferromagnetic metal in contrast to the case of Co-NW. The Co-FEBID nanowire has a nanogranular microstructure. In general, structures prepared by FEBID belong to the class of disordered electronic materials with different degrees of disorder, ranging from a low impurity concentration in a well-ordered polycrystalline structure to strongly disordered amorphous materials in the opposite limiting case. With regard to the microstructural and electrical transport properties, the Co-FEBID nanowire is located between these limiting cases. As inclusions of carbon and oxygen have been observed by EDX spectroscopy, see also table~\ref{tEDX}, we argue that C and O are impurities responsible for suppressing the proximity effect, since these elements represent effective scattering centers hindering Cooper pairs to spread through the fine-dispersed deposit. On the microscopical level, their role in the scattering process is discussed next.

Consider a region in Co-FEBID which forms one magnetic domain and hence all spins in this region point in one direction. Then, to survive in this environment, a Cooper pair should also have spins pointing in the same direction. Adding a scattering process at the domain boundaries results in changes of the orbital momentum and, due to the spin-orbital interaction, in changes of the orientation of spins. When the spins start to flip-over this leads to the pair-breaking effect. This scenario strongly depends on the symmetry of both, the scattering center and the wave function of the propagating Cooper pair. In our case, Cooper pairs are induced from the W floating electrode having $s$-wave symmetry, while cobalt has $d$-wave symmetry, oxygen has $p$-wave valence orbitals, and carbon has $s$-$p$-hybride orbitals. Both oxygen and carbon have small atom radii so that both elements can be incorporated into interstitials of the Co hcp lattice. We theqrefore assume that strong pair-breaking will result in $p$-wave like scatters for Cooper pairs in an odd-frequency spin-triplet state. This would efficiently suppress a long-range proximity effect in Co-FEBID.

\section{Quantification of proximity length}\label{sXi}

In this subsection, we analyze the resistance drops in the vicinity of $T_c$ of the W inducer electrode and quantify the proximity lengths for all nanowires.
\begin{figure}
    \centering
    \includegraphics[width=0.3\textwidth]{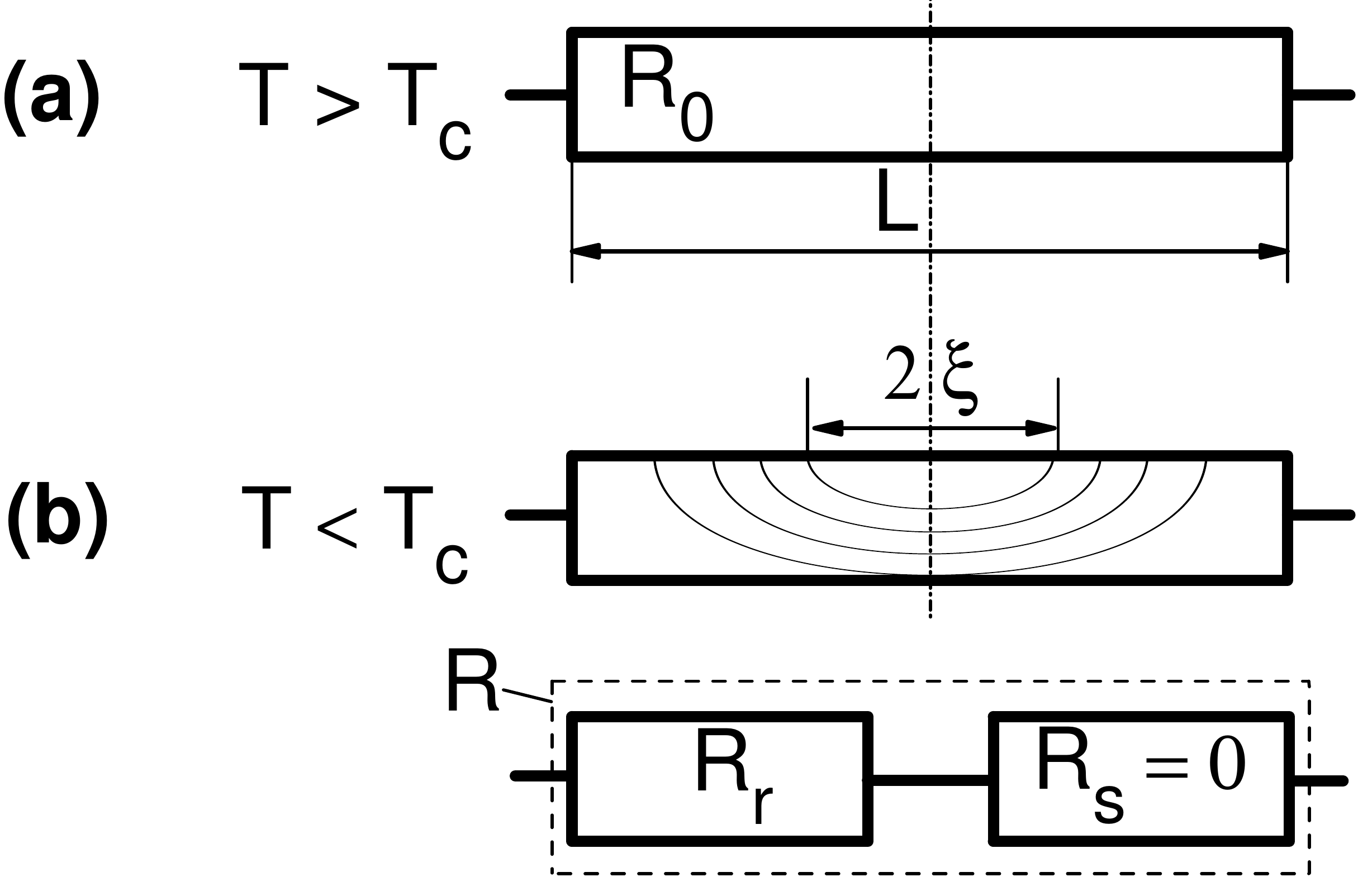}
    \caption[]
    {(a) At $T > T_c$ the nanowire is modeled as resistor $R_0$. (b) At $T < T_c$ Cooper pairs propagate from the superconducting inducer into the nanowire and a finite fraction of the nanowire becomes superconducting. The spatial extent of the superconducting condensate in the nanowire at different temperatures is shown by the contour lines. Bottom panel: The model electrical circuit used for the quantification of the proximity length in the nanowires.}
   \label{fRvL}
\end{figure}

Our treatment of the resistance data in the 8-probe geometry relies upon the model electrical circuit sketched in figure~\ref{fRvL}. Consider a nanowire in contact with a superconducting inducer electrode located at the middle of the nanowire. Assume that the nanowire of length $L$ has an ideal cylindrical shape. Furthermore, assume that the current distribution in the cross-section of the nanowire is homogenous and it is not affected by the Pt-FIBID voltage electrodes and the W-FIBID inducer. Then, at $T > T_c$ the nanowire can be regarded as a resistor with the normal-state resistance $R_0$, refer to figure~\ref{fRvL}(a). By contrast, at $T < T_c$, when Cooper pairs start to propagate from the superconducting inducer into the nanowire, a finite fraction of the nanowire becomes superconducting, see figure~\ref{fRvL}(b). As the superconducting proximity length depends on temperature, each contour in the figure corresponds to different $T$. Accordingly, the respective parts of the nanowire within the semispheres have zero resistance. Obviously, the length of the superconducting fraction is twice the proximity length $2\xi$, since Cooper pairs propagate in both directions along the nanowire axis. The remaining part of the circuit of length $L-2\xi$ is in the normal state with the residual resistance $R_r$ = $R_0({L-2\xi})/L$. The effective resistor model is represented in the bottom panel of figure~\ref{fRvL}(b). According to this model circuit, the total resistance of the measured nanowire section is
\begin{equation}
   \label{xi01}
   R = R_0({L-2\xi})/L + R_s({2\xi/L}),
\end{equation}
from which the proximity length is
\begin{equation}
  \label{xi02}
  \xi={L(R_0-R)/2R_0}.
\end{equation}

\begin{figure*}
\centering
    \includegraphics[width=0.7\textwidth]{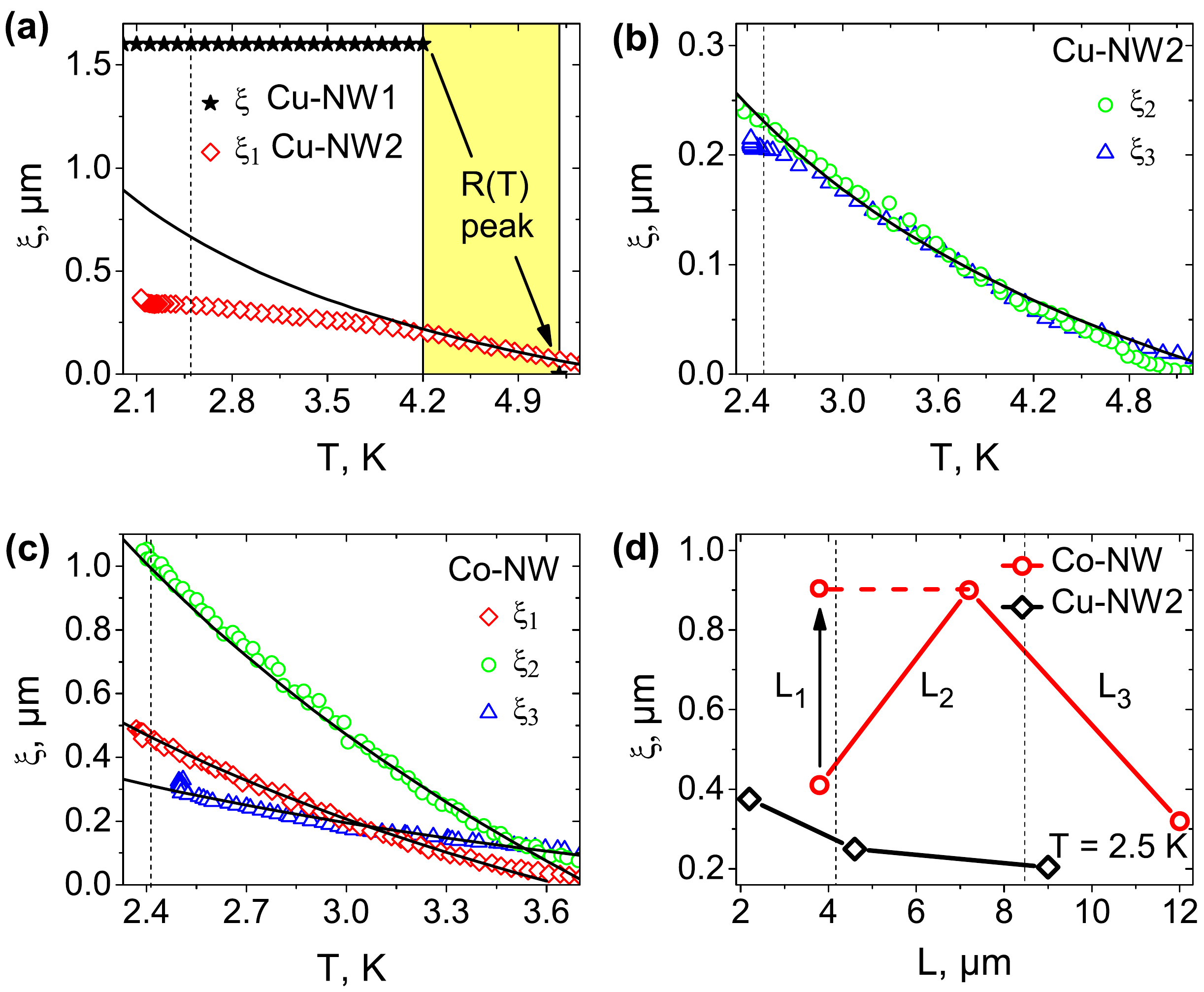}
    \caption[]
    {The temperature dependences of the proximity lengths for Cu-NW1, Cu-NW2~(a) and (b), and Co-NW~(c), respectively. $\xi_1$, $\xi_2$ and $\xi_3$ correspond to the distances between the voltage leads $L_1$, $L_2$ and $L_3$, respectively. The solid lines are fits to an expression of the form $\xi(T) \propto \sqrt{1/T}$. Note a factor of two deviation of $\xi(T)$ from the fit curve in (a) for $T < 4$~K which we attribute to the enhanced degree of disorder caused by the processing by FIBID. (d)~The proximity length deduced for the different measured sections for the same samples at $2.5$~K shown by the vertical lines in (a)-(c). The vertical axis and the dashed line in (d) point out that the experimentally measured value of the proximity length for the $L_1$ section in Co-NW is likely underestimated. Refer to text for details.}
   \label{fXivT}
\end{figure*}

An analogous model can be applied to Cu-NW1 in the 4-probe geometry where Cooper pairs spread from the superconducting voltage electrodes inwards the nanowire. In this way, using equation~\eqref{xi02} and the experimental data reported in Figs.~\ref{fCuNW1RvT}(b), \ref{fCuNW2RvT}, \ref{fCoNWRvT}(b), and \ref{fCoEBID}(b), the proximity lengths for all the samples have been calculated. The central results of these calculations are the $\xi(T)$ curves for Cu-NW2 and Co-NW shown in figures~\ref{fXivT}(a), (b), and (c) which we consider now in more detail.

Before entering the discussion it should be noted that in the reference sample Cu-NW1 the proximity length at 2.5~K ($\approx T_c/2$ of the inducer electrode) is $\xi=1.6~\mu$m [figure~\ref{fXivT}(a)]. This is an exemplary value for the ``classical'' proximity length in pure diamagnetic materials. It should be noted that in the vicinity of $T_c$ we did not succeed in quantifying $\xi(T)$ as the exact shape of the resistance drop is masked by the resistance peak. At 2.5~K, the proximity length in Cu-NW2 is a factor of five shorter and we attribute this to the degradation of the conducting properties of Cu-NW in the regions under the impact of the ion beam. This results in the deviation of $\xi(T)$ from the fit curve in figure~\ref{fXivT}(a). The suppression of the proximity length is believed to be caused by the enhanced scattering of the Cooper pairs in the defect-rich regions of the nanowire. Interestingly, the calculated proximity length for Co-NW at 2.5~K is $0.5 - 1~\mu$m, attesting to an even more long-ranged effect as compared to Cu-NW2. By contrast, the calculated proximity length for Co-FEBID at $2.5$~K is of the order of $100$~nm (not shown). According to our analysis of the microstructure and the scattering mechanisms in this sample in section~\ref{sResults}, we believe that the calculated value is not related to the proximity length but rather to the length of the inducer short-circuited nanowire section, as its doubled value is very close to the width of the W-FIBID inducer electrode. That is, in the case of the nanogranular Co-FEBID structure we have not been able to reliably observe the proximity effect, due to the spatial resolution limitations mediated by the width of the superconducting inducer electrode. Turning back to the $\xi(T)$ dependences for Cu-NW2 and Co-NW, these can be fitted well to an expression of the form $\xi(T) \propto \sqrt{1/T}$. This is in good agreement with the theoretical predictions~\cite{Buz05rmp} for the temperature dependence of the superconducting proximity length in the diffusive limit.

\begin{figure}
\centering
    \includegraphics[width=0.45\textwidth]{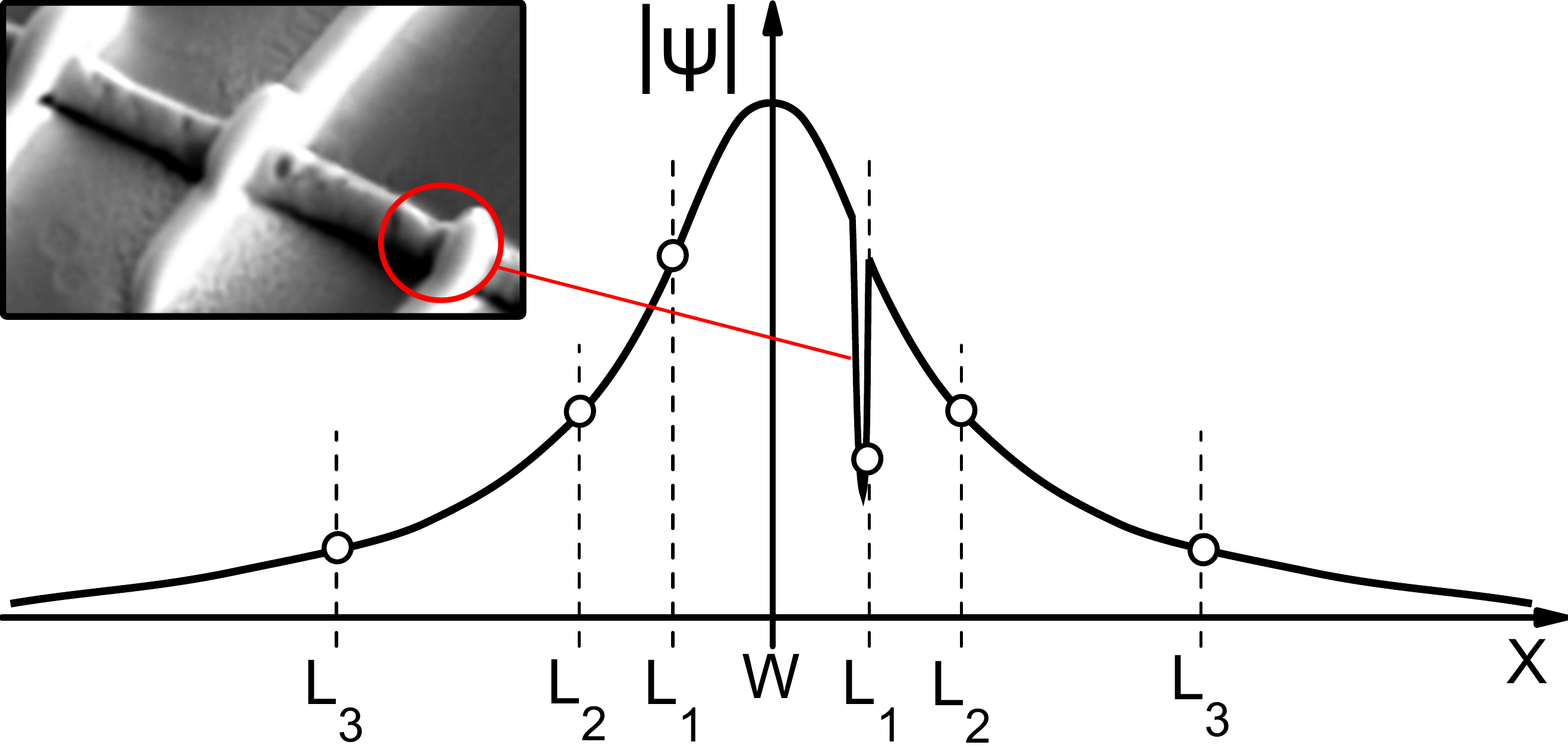}
    \caption[]
    {Interruption of the order parameter by a grain boundary. Inset: SEM image of the grain boundary in the vicinity of one of the $V_1$ voltage leads.}
   \label{OrderParametr}
\end{figure}

Figure~\ref{fXivT}(d) illustrates the calculated proximity length in Cu-NW2 and Co-NW for the different measured sections at $2.5$~K. One can see from the data that $\xi(L)$ for Cu-NW decreases with increasing distance between the voltage leads. We attribute this to the effectively increasing scattering of the Cooper pairs in the disorder-rich regions underneath the contacts as the number of contacts between the voltage leads rises. Surprisingly, the dependence $\xi(L)$ for Co-NW is not monotonic in distance and we decided to undertake a post-measurement inspection of the contact regions in the SEM. Under the microscope, we observed that the location of one of the $V_1$ potential probes is very close to a grain boundary between neighboring crystallites (see the inset of figure~\ref{OrderParametr}). As the superconducting triplet order parameter is coupled to the ferromagnetic ordering, the latter aligns the Cooper pair spins along the magnetization direction. As the direction of magnetization of neighboring domains will in general differ, the superconducting order parameter is suppressed at the grain boundary with respect to its intra-domain value, see also sketch in figure~\ref{OrderParametr}. Accordingly, if a contact lead is placed in the vicinity of the grain boundary, the maximal amplitude of the superconducting order parameter can not be probed. This, in turn, results in a reduction of the relative resistance drop due to the proximity effect. This is why we believe that the calculated proximity length for the inner voltage leads is underestimated and in fact it can be as large as $1~\mu$m.

\section{Conclusion}\label{sConclusion}

We have experimentally studied proximity effect-induced superconductivity by electrical resistance measurements in single metallic and ferromagnetic nanowires. Specifically, four different samples have been investigated. These were two single-crystal high-quality Cu nanowires with a close-to-bulk resistivity, one polycrystalline Co nanowire, and one rectangular nanowire-shaped fine-dispersed nanogranular Co structure. The different microstructural properties of the samples allowed us to investigate qualitatively different cases of proximity-induced superconductivity coexisting with other effects. In particular, we identified and quantified a large resistance contribution of the ion-beam damaged regions in the case of high-quality Cu nanowires, a localization-like low-temperature transport in the polycrystalline Co nanowire owing to a large resistivity of the grain boundaries, and strong pair-breaking effects due to the wave function symmetry-altering scattering in the fine-dispersed nanogranular Co structure. In all the cases, proximity-induced superconductivity became apparent via resistance drops just below the transition temperature of the W superconducting inducer electrode ($\approx5.1$~K). By using simple resistance models we succeeded in accounting for the resistance contributions stemming from the ion-beam damaged nanowire regions and to quantify the proximity lengths as a function of temperature in all the samples. Our key observation is that in the polycrystalline Co nanowire the observed effect is long-ranged, with a proximity length of the order of $1~\mu$m at 2.5~K. Moreover, this long-ranged effect is unsusceptible to magnetic fields up to $11$~T being limited only by the critical field of the superconducting electrode. All this attests to the spin-triplet nature of the observed proximity effect in cobalt. Interestingly, the same effect has not been observed in the nanogranular Co nanowire structure. The performed microstructural analysis of this sample has allowed us to explain the enhanced pair-breaking effects by the wave function symmetry-altering scattering at the boundaries of nano-grains. Quantitatively, the temperature dependences of the superconducting proximity length in the single-crystal Cu and the polycrystalline Co could be fitted very well to an expression of the form $\xi(T) \propto \sqrt{1/T}$ in a wide temperature range. Finally, one remark is in order concerning the calculated proximity length and the dimensions of crystalline grains in the polycrystalline Co nanowire. For the temperature range where $\xi$ is larger than the typical dimension of the grains question remains, if the propagation of a Cooper pair is limited by the grain size or if paired electrons can tunnel through the grain boundary maintaining the phase coherence.

\section{Acknowledgments}
The authors thank R. Sachser for support in automating the data acquisition. Discussions with K. Arutyunov and A. Buzdin are acknowledged. This work was supported by the Beilstein Institut, Frankfurt/M, within the research collaboration NanoBiC.

\bibliography{./proximity}
\bibliographystyle{aipnum4-1}

\end{document}